\documentclass[notitlepage,prd,preprintnumbers,longbibliography,showpacs]{revtex4-2}
\usepackage[colorlinks,linkcolor=magenta,anchorcolor=cyan,citecolor=blue]{hyperref}
\usepackage{slashed,color,amsmath,amssymb,mathrsfs}
\usepackage[utf8]{inputenc}
\usepackage{graphicx}
\usepackage{longtable}
\usepackage{array}
\usepackage{autobreak}
\usepackage{booktabs}
\usepackage{cancel}
\usepackage{bm}
\usepackage{physics}
\usepackage{accents}

\allowdisplaybreaks

\newcommand{\la}{\langle}
\newcommand{\ra}{\rangle}

\newcommand{\fp}{f_+}

\newcommand{\fm}{f_-}

\newcommand{\chip}{\chi_+}

\newcommand{\chim}{\chi_-}

\newcommand{\ahe}{{\hat{\alpha}_\perp}}

\newcommand{\ahp}{{\hat{\alpha}_\parallel}}

\newcommand{\As}{\hat{\mathscr{A}}}

\newcommand{\Hs}{\hat{\mathscr{H}}}

\newcommand{\Vs}{\hat{\mathscr{V}}}

\newcommand{\Vt}{V}

\newcommand{\chiph}{\hat{\chi}_+}

\newcommand{\chimh}{\hat{\chi}_-}

\begin{document}

{
\title{Higher-order chiral Lagrangians with vector meson nonet in different representations}
\author{Wei Guo$^{1,2}$}
\email{wguo@cwnu.edu.cn}
\author{Qin-He Yang$^{1,3}$}
\author{Shao-Zhou Jiang$^{1}$}
\email[Corresponding author.]{jsz@gxu.edu.cn}
\affiliation{$^{1}$Guangxi Key Laboratory for Relativistic Astrophysics, School of Physical Science and Technology, Guangxi University, Nanning 530004, China\\
$^{2}$School of Physics and Astronomy, China West Normal University, Nanchong 637009, China\\
$^{3}$School of Physics and Electronics, Hunan University, Changsha 410082, China}
\date{\today}

\begin{abstract}

In this paper, chiral Lagrangians with vector meson nonet are constructed across multiple representations, including those to the next-to-leading order in the vector-field representation, as well as to the next-to-next-to-leading order in the tensor-field and hidden local symmetry representations. For the next-to-leading order octet, redundant terms in the other literature are also identified in both the tensor-field and hidden local symmetry representations. Additionally, the equivalence between the tensor-field and hidden local symmetry representations is examined.
\end{abstract}
\maketitle

\section{Introduction}
Quantum chromodynamics (QCD) is the fundamental theory describing strong interactions. However, due to non-perturbative effects, QCD often cannot provide precise predictions in the low-energy region. To study low-energy QCD involving pseudoscalar mesons, one can use an effective theory at the hadronic level, leveraging chiral symmetry and its spontaneous breaking. This approach is known as chiral perturbation theory (ChPT) \cite{callan_structure_1969,coleman_structure_1969,weinberg_phenomenological_1979,gasser_chiral_1984,gasser_chiral_1985}. ChPT serves as an effective field theory of QCD at low energies. In ChPT, the chiral Lagrangian encompasses all effective interactions that respect Lorentz invariance, chiral symmetry, charge conjugation, and parity. Rather than an expansion in terms of the large strong coupling constant, ChPT is formulated as an expansion in the typical momentum scale ($p$). According to Weinberg's power-counting rule, tree-level and loop diagrams contribute terms proportional to powers of $p/\Lambda$, where $\Lambda$ represents the chiral symmetry-breaking scale. Thus, calculating a physical quantity with a specific accuracy requires only a finite number of interactions. Currently, pseudoscalar mesonic chiral Lagrangians have been constructed up to $\mathcal{O}(p^8)$ in both $SU(2)$ and $SU(3)$ symmetries \cite{gasser_chiral_1984,gasser_chiral_1985,fearing_extension_1996,bijnens_mesonic_1999,bijnens_anomalous_2002,bijnens_order_2018,Bijnens:2023hyv}.

ChPT can be extended to higher-energy ranges to incorporate additional particles, such as vector mesons, which exhibit unique properties. With masses that fall between those of pseudoscalar mesons and baryons, vector mesons present challenges for consistent power counting\cite{leupold_towards_2012,Park:2024mrw}. In the literature, vector mesons are sometimes treated as light degrees of freedom \cite{prades_massive_1994,Rosell:2004mn,Rosell:2006dt,lutz_radiative_2008,leupold_hadronic_2009,Rosell:2009yb,kampf_renormalization_2010,Pich:2010sm,danilkin_causality_2011,terschlusen_electromagnetic_2012,danilkin_photon-fusion_2013,gao_parton_2014,guo_light_2019,Braghin:2021qmu,Pelaez:2025wma,Geng:2024oaj}, while other studies consider them as heavy degrees of freedom \cite{jenkins_chiral_1995,Bijnens:1996nq,bijnens_chiral_1998,djukanovic_path_2010,bruns_chiral_2013,Mantysaari:2022kdm}.

There are several representations used to describe vector mesons: The traditional vector-field (or matter-field/Proca field) representation \cite{weinberg_nonlinear_1968,ecker_chiral_1989}, the antisymmetric tensor-field representation \cite{gasser_chiral_1984,ecker_chiral_1989,ecker_role_1989}, the hidden local symmetry (HLS) representation \cite{bando_is_1985,bando_vector_1985,bando_composite_1985,fujiwara_non-abelian_1985,tanabashi_chiral_1993,tanabashi_formulations_1996,harada_wilsonian_2001,harada_hidden_2003,ma_hidden_2005}, and the massive Yang-Mills representation \cite{Schwinger:1967tc,Wess:1967jq,Gasiorowicz:1969kn,Kaymakcalan:1983qq,Meissner:1987ge}. This work focuses on the first three, as the massive Yang-Mills representation is seldom used today.

While distinct, these representations are interrelated. The vector-field representation, the most traditional, has its connections to other representations thoroughly documented in Refs. \cite{ecker_chiral_1989,bijnens_tensor_1996,birse_effective_1996,harada_hidden_2003,kampf_different_2007}. In the tensor-field representation, vector mesons are represented by antisymmetric tensors, which transform homogeneously under chiral symmetry \cite{ecker_chiral_1989,ecker_role_1989,bijnens_tensor_1996,kampf_different_2007}. The vector-field and tensor-field representations share many similarities, with some studies treating them together \cite{ecker_chiral_1989,bijnens_tensor_1996,kampf_different_2007}. The tensor-field representation offers certain advantages: It provides a straightforward generalization of the Dirac equation for spin-1/2 particles to spin-1 particles and corresponds to the $(1,0) \oplus (0,1)$ representation of the Lorentz group, naturally encompassing two spin-1 degrees of freedom among its six degrees of freedom. In contrast, the vector-field representation uses the $(1/2,1/2)$ Lorentz representation, requiring an auxiliary condition to eliminate the unwanted spin-0 component. The tensor-field Lagrangian is also simpler in form than that of the vector-field representation. Further details on the relations between these two representations can be found in Refs. \cite{ecker_chiral_1989,birse_effective_1996,bijnens_tensor_1996,tanabashi_formulations_1996,kampf_different_2007}.

In the HLS representation, an artificial local symmetry is introduced into the nonlinear sigma model through the choice of specific field variables, which can be removed by selecting a gauge, such as the unitary gauge. This gauge symmetry is not a true symmetry but a redundancy in the description. The HLS representation's connections with other representations are discussed in Refs. \cite{ecker_chiral_1989,pallante_anomalous_1993,tanabashi_formulations_1996,harada_hidden_2003}. The massive Yang-Mills representation is a particular gauge choice within the generalized HLS model, determined by specific parameter values, with further relations to other representations elaborated in Refs. \cite{Meissner:1986tc,Yamawaki:1986zz,Golterman:1986cz,ecker_chiral_1989}.

Two types of equivalence relations exist among these representations \cite{pallante_anomalous_1993,kampf_different_2007}. The first type indicates that different representations share the same structure without integrating out vector mesons \cite{tanabashi_formulations_1996,harada_hidden_2003}, while the second type arises when different representations yield equivalent results after vector fields are integrated out \cite{ecker_chiral_1989,bijnens_tensor_1996,kampf_different_2007}. The first type of equivalence applies at the level of resonance Lagrangians, whereas the second applies at the level of pseudoscalar mesonic chiral Lagrangians. Importantly, the first-type equivalences imply the second-type, but the reverse is not necessarily true.

After integrating out the vector fields and comparing with pseudoscalar mesonic chiral Lagrangians \cite{gasser_chiral_1984,gasser_chiral_1985,bijnens_chiral_2007,bijnens_mesonic_1999}, the equivalence of these representations at the level of pseudoscalar mesonic chiral Lagrangians becomes evident \cite{ecker_chiral_1989,bijnens_tensor_1996,kampf_different_2007}. A recent approach, the first-order formalism, enables calculation of resonance contributions to low-energy constants (LECs) in ChPT at next-to-next-to-leading order (NNLO) by integrating out vector fields \cite{kampf_different_2007}. This method underscores the equivalence between the vector-field and tensor-field representations at the NNLO level.

Currently, although there is extensive research on ChPT with vector mesons, most studies focus on treating vector mesons as resonance states to investigate the properties of pseudoscalar mesons \cite{RuizFemenia:2003hm,Cirigliano:2004ue,Cirigliano:2005xn,cirigliano_towards_2006,Dai:2019lmj,Kampf:2006bn,Kampf:2011ty}. For such purposes, the lower-order chiral Lagrangian is usually sufficient. However, some studies concentrate on vector mesons themselves, analyzing aspects such as their masses \cite{Bijnens:1996nq,Bijnens:1997ni,bruns_infrared_2005,Djukanovic:2009zn,Bruns:2013tja,Kawaguchi:2015gpt,Bavontaweepanya:2018yds}, electromagnetic properties \cite{danilkin_photon-fusion_2013,Unal:2019eum}, decay constants \cite{Bijnens:1998di,Geng:2024oaj}, radiative decays \cite{lutz_radiative_2008,Lange:2021lxb}, and scattering with other particles at high energies \cite{ma_hidden_2005,Zhou:2014ila,terschlusen_electromagnetic_2012,Djukanovic:2018pep,Djukanovic:2004mm}. Research on vector mesons at the next-to-leading order (NLO) has also been conducted, and studies at the NNLO are emerging \cite{cirigliano_towards_2006,Kampf:2006bn,Kampf:2011ty,Dai:2019lmj,Kawaguchi:2015gpt}.

To date, vector mesonic chiral Lagrangians have been developed up to NLO in the tensor-field and HLS representations \cite{bando_is_1985,ecker_chiral_1989,tanabashi_chiral_1993,cirigliano_towards_2006}. However, for studies focused on vector mesons themselves, lower-order Lagrangians may not offer the required accuracy. In this paper, we systematically construct chiral Lagrangians for the vector-field, tensor-field, and HLS representations, extending to NNLO and covering both the normal and anomalous cases, and the strong equivalence between the tensor-field and HLS representations is also analyzed. In addition, in order to expand the application scope of the chiral Lagrangian, the singlet vector meson is also considered, i.e., the chiral Lagrangian is expanded to the vector meson nonet. This Lagrangian can be applied not only to low-energy vector mesons but also to heavy-vector-meson fields. For example, treating $J/\Psi$ as a vector meson singlet, it can be used to study the interaction between $J/\Psi$ and pseudoscalar mesons\cite{Chen:2014yta,Yan:2023nqz,Zhang:2025vbf}.

While multi-vector-meson interactions have been explored in earlier research, the complexity of these interactions in the vector-field and tensor-field representations has often restricted studies to single-vector-meson vertices. Consequently, our analysis concentrates on single-vector-meson vertices within these representations. Although higher-order chiral Lagrangians contain numerous terms, only a few play a significant role in the specific physical phenomena. Advances in technology and the development of specialized software and computer algebra systems now enable efficient handling of the computational challenges posed by these complex terms.

This paper focuses on a tree-level analysis. Extending the discussion to include loop diagrams poses significant challenges, as the presence of vector mesons disrupts the conventional power-counting rule \cite{bruns_infrared_2005}. Specifically, loop diagrams introduce large imaginary components that violate power counting, and these terms cannot be absorbed using standard counterterms or renormalized parameters within the traditional renormalization scheme. Currently, the only viable solution to this problem is the complex-mass scheme \cite{Denner:1999gp,Denner:2005fg,Denner:2006ic,Actis:2006rc,Actis:2008uh,Djukanovic:2009zn,Djukanovic:2009gt,Denner:2014zga,Djukanovic:2015gna,Gegelia:2016xcw,Gegelia:2016pjm}. In this scheme, the masses and widths of vector mesons are assigned power-counting orders of $\mathcal{O}(p^0)$ and $\mathcal{O}(p^1)$, respectively. The power-counting rule for a vector meson propagator depends on the external momenta: It is counted as $\mathcal{O}(p^0)$ if it carries large external momentum and $\mathcal{O}(p^1)$ otherwise. Additionally, derivatives acting on vector mesons are assigned an order of $\mathcal{O}(p^0)$. Meanwhile, the power counting for pseudoscalar mesons remains consistent with standard pseudoscalar mesonic ChPT. In the complex-mass scheme, power-counting violations can be controlled by introducing additional counterterms, albeit with more complex parameter structures. This approach preserves the power-counting rule, allowing for the application of the chiral expansion even at the loop level.

This paper is structured as follows. Sec. \ref{sec:chiral_su_n} provides a brief overview of pseudoscalar and vector mesons in ChPT. Secs. \ref{vfa} to \ref{ha} detail the methods for constructing chiral Lagrangians with vector mesons in the vector-field, tensor-field, and HLS representations, respectively. Sec. \ref{rcl} outlines the resulting chiral Lagrangians, while Sec. \ref{sec:equivalence} examines the equivalence between the HLS and tensor-field representations. Finally, Sec. \ref{sec:conclusions} gives a summary and some comments.

\section{Mesons in chiral perturbation theory} \label{sec:chiral_su_n}
In QCD, the Lagrangian with external sources is
\begin{align}
\mathscr{L}_{\mathrm{QCD}} = \mathscr{L}_{\mathrm{QCD}}^{0} + \bar{q} \gamma_\mu (v^\mu + a^\mu \gamma_5) q - \bar{q} (s - i p \gamma_5) q,
\end{align}
where $\mathscr{L}_{\mathrm{QCD}}^{0}$ is the original QCD Lagrangian, and $q$ denotes the light quark fields. Here, $v^{\mu}$, $a^{\mu}$, $s$, and $p$ represent the vector, axial-vector, scalar, and pseudoscalar sources, respectively. For simplicity, the tensor source and the $\theta$ term are omitted. To facilitate comparison with the pseudoscalar mesonic chiral Lagrangian, $a^{\mu}$ is considered traceless, and $v^{\mu}$ is traceable only in the two-flavor anomalous part; in other contexts, it is taken to be traceless.

If the masses of the light quarks are ignored, the QCD Lagrangian possesses a global $SU(N_f)_L \times SU(N_f)_R$ chiral symmetry, where $N_f = 2$ or $3$. This symmetry spontaneously breaks to an $SU(N_f)_V$ symmetry, resulting in the emergence of pseudoscalar Goldstone bosons associated with the symmetry breaking. These bosons correspond to the lowest pseudoscalar mesons, represented by $U$. Under a chiral transformation, $U$ transforms as $U \rightarrow g_L U g_R^{\dagger}$, where $g_R$ and $g_L$ are elements of the $SU(N_f)_L$ and $SU(N_f)_R$ chiral groups, respectively.

For convenience, an auxiliary field $u$ is introduced to describe the pseudoscalar meson fields and $u^2 = U$\cite{Scherer:2012xha,Petrov:2016azi},
\begin{align}
U(x)=\exp(\frac{i\sqrt{2}}{F_0}\pi(x)),
\end{align}
where $F_0$ is just a low-energy constant, and $\pi(x)$ represent pseudoscalar meson fields,
\begin{align}
\pi(x)=\sum_{i=1}^{8}\frac{1}{\sqrt{2}}\lambda_i\pi_i(x)=\left(\begin{array}{ccc}
\frac{\pi^{0}}{\sqrt{2}}+\frac{\eta_{8}}{\sqrt{6}} &\pi^{+} & K^{+} \\
\pi^{-} & -\frac{\pi^{0}}{\sqrt{2}}+\frac{\eta_{8}}{\sqrt{6}} & K^{0} \\
 K^{-} & \bar{K}^{0} & -\frac{2 \eta_{8}}{\sqrt{6}}
\end{array}\right),
\end{align}
where $\lambda_i$ are Gell-Mann matrices. The chiral transformation of the pseudoscalar meson fields $\pi(x)$ here is nonlinear, so it is often also called a nonlinear realization. Under a chiral transformation, $u$ transforms as $u \rightarrow g_L u h^{\dagger} = h u g_R^{\dagger}$, where $h$ is a compensator field that depends on the meson fields. This transformation property of $u$ simplifies the construction of chiral Lagrangians, and we will adopt it in this paper. For further details on ChPT, readers may consult Refs.~\cite{Scherer:2012xha, Burgess:2020tbq, Petrov:2016azi} and references therein.

The vector meson nonet is represented by a $3\times 3$ matrix in the vector-field representation
\begin{align}
V^{\mu}=\left( \begin{array}{ccc}{\displaystyle\frac{\rho^{0}}{\sqrt{2}}+\frac{\omega_8}{\sqrt{6}}+\frac{\omega_0}{\sqrt{3}}  } & {\rho^{+}} & {K^{*+}} \\
{\rho^{-}} & \displaystyle{-\frac{\rho^{0}}{\sqrt{2}}+\frac{\omega_8}{\sqrt{6}} +\frac{\omega_0}{\sqrt{3}}} & {K^{* 0}} \\
{K^{*-}} & {\overline{K}^{* 0}} & \displaystyle{-\frac{2 \omega_8}{\sqrt{6}}}+\frac{\omega_0}{\sqrt{3}}
\end{array}\right)^{\mu}.\label{dV}
\end{align}
The vector field $V^\mu$ transforms as $h V^{\mu} h^{\dagger}$ under chiral rotations. In the tensor-field representation, $V^\mu$ is replaced by an antisymmetric tensor $W^{\mu\nu}$, where each element in $V^{\mu}$ corresponds to an antisymmetric tensor field. The HLS representation is more complex and will be discussed in detail in Sec. \ref{ha}.

To account for the finer effects such as $\rho^0-\omega$ mixing \cite{Urech:1995ry} or $\omega-\phi$ mixing \cite{Kucukarslan:2006wk}, $\rho^0$, $\omega_8$ and $\omega_0$ should be included in the mixing. For the ideal mixing,
\begin{align}
\begin{pmatrix}
\omega^\mu\\
\phi^\mu
\end{pmatrix}
=\begin{pmatrix}
\sqrt{\frac{1}{3}} & \sqrt{\frac{2}{3}}\\
-\sqrt{\frac{2}{3}} & \sqrt{\frac{1}{3}}
\end{pmatrix}
\begin{pmatrix}
\omega_8^\mu\\
\omega_0^\mu
\end{pmatrix}.
\end{align}

This paper does not assume isospin symmetry; if one considers isospin breaking, such as the mass difference between $\rho^0$ and $\rho^\pm$, then $\rho^0$, $\omega$, and $\phi$ mixing needs to be included, and $s$ should be replaced by $\mathrm{diag}(m_u,m_d,m_s)$ in the calculation \cite{Bijnens:1996nq,Bijnens:1997ni}.

The following three sections discuss the construction of chiral Lagrangians with vector mesons in different representations. In this paper, we restrict our attention to single-vector-meson vertices within the vector-field and tensor-field representations.

\section{Vector-field representation}\label{vfa}
\subsection{Building blocks and transformation properties}\label{bbp}
In addition to the vector meson fields, the building blocks in the vector-field representation are identical to those in the pseudoscalar mesonic chiral Lagrangians \cite{bijnens_mesonic_1999,bijnens_chiral_2007},
\begin{align}
u^{ \mu } & = i \left\{ u ^ { \dagger } \left( \partial^{ \mu } - i r^{ \mu } \right) u - u \left( \partial^{ \mu } - i \ell^{ \mu } \right) u ^ { \dagger } \right\}, \notag\\
\chi _ { \pm } & = u ^ { \dagger } \chi u ^ { \dagger } \pm u \chi ^ { \dagger } u, \notag\\
f_{+}^{\mu \nu} &=u F_{L}^{\mu \nu} u^{\dagger} + u^{\dagger} F_{R}^{\mu \nu} u,\notag\\
f_{-}^{\mu \nu} &=u F_{L}^{\mu \nu} u^{\dagger} - u^{\dagger} F_{R}^{\mu \nu} u=-\nabla^{\mu} u^{\nu}+\nabla^{\nu} u^{\mu} ,\notag\\
h^{\mu \nu} &=\nabla^{\mu} u^{\nu}+\nabla^{\nu} u^{\mu},\label{bb}
\end{align}
where $r^{\mu}=v^{\mu}+a^{\mu}$, $\ell^{\mu}=v^{\mu}-a^{\mu}$, $\chi=2 B_0(s+i p)$, $F_{R}^{\mu \nu}=\partial^{\mu} r^{\nu}-\partial^{\nu} r^{\mu}-i\left[r^{\mu}, r^{\nu}\right]$, $F_{L}^{\mu \nu}=\partial^{\mu} \ell^{\nu}-\partial^{\nu} \ell^{\mu}-i\left[\ell^{\mu}, \ell^{\nu}\right]$, and $B_0$ is a constant related to the quark condensate. These building blocks exhibit the same chiral rotation properties as $V^{\mu}$, i.e., $X \rightarrow h X h^\dagger$, where $X$ represents any building block. The covariant derivative is defined by
\begin{align}
\nabla^{\mu} X=\partial^{\mu} X+\left[\Gamma^{\mu}, X\right],
\end{align}
where $\Gamma_{\mu}$ is the chiral connection,
\begin{align}
\Gamma^{\mu}=\frac{1}{2}\left\{u^{\dagger}\left(\partial^{\mu}-i r^{\mu}\right) u+u\left(\partial^{\mu}-i \ell^{\mu}\right) u^{\dagger}\right\}.
\end{align}

To construct the vector mesonic chiral Lagrangians, $V^\mu$, $X$, and their derivatives are considered. Some literature also introduces a building block $\hat{V}^{\mu\nu}=\nabla^{\mu}V^{\nu}-\nabla^{\nu}V^{\mu}$ \cite{ecker_chiral_1989,kampf_different_2007}. However, as only single-vector-meson vertices are considered, the covariant derivatives acting on $V^{\mu}$ can be transferred to other building blocks through partial integration, which will be discussed in Eq. \eqref{pir}. Therefore, aside from the kinetic term, we do not include this building block.

In addition to chiral symmetry, chiral Lagrangians also need to be invariant under parity ($P$), charge conjugation ($C$), and Hermitian conjugation (h.c.) transformations. The transformation properties of these building blocks have been discussed extensively in the literature \cite{bijnens_mesonic_1999,bijnens_chiral_2007,ecker_chiral_1989,ecker_role_1989}. ChPT is usually expanded in powers of the momentum ($p$) of the external source, denoted as $\mathcal{O}(p^n)$, where $n$ is referred to as the chiral dimension (abbreviated as Dim). Table \ref{blbt} summarizes these properties for reference.

\begin{table*}[!h]
\caption{\label{blbt}Chiral dimension (Dim), parity ($P$), charge conjugation ($C$) and hermiticity (h.c.) of the building blocks in the vector-field and tensor-field representations.}
\begin{center}
\begin{tabular}{ccccc}
\hline\hline
& Dim &                $P$                 &                $C$                &               h.c.                \\
\hline
$u^{\mu}$             &  1  &             $-u_{\mu}$             &           $(u^{\mu})^T$           &             $u^{\mu}$             \\
$h^{\mu\nu}$            &  2  &           $-h_{\mu\nu}$            &         $(h^{\mu\nu})^T$          &           $h^{\mu\nu}$            \\
$\chi_{\pm}$            &  2  &          $\pm\chi_{\pm}$           &         $(\chi_{\pm})^T$          &         $\pm \chi_{\pm}$          \\
$f_{\pm}^{\mu\nu}$         &  2  &        $\pm f_{\pm\mu\nu}$         &    $\mp (f_{\pm}^{\mu\nu})^T$     &        $ f_{\pm}^{\mu\nu}$        \\
$V^{\mu}$             &  0  &                $V_{\mu}$                &           $-(V^{\mu})^T$            &             $V^{\mu}$              \\
$W^{\mu\nu}$             &  0  &                $W_{\mu\nu}$                &           $-(W^{\mu\nu})^T$            &             $W^{\mu\nu}$              \\
$\varepsilon^{\mu\nu\lambda\rho}$ &  0  & $-\varepsilon_{\mu\nu\lambda\rho}$ & $\varepsilon^{\mu\nu\lambda\rho}$ & $\varepsilon^{\mu\nu\lambda\rho}$ \\
\hline\hline
\end{tabular}
\end{center}
\end{table*}

The vector meson masses are about \cite{ParticleDataGroup:2024cfk}
\begin{align}
\rho^\pm, \rho^0(775 ~\mathrm{MeV}),\quad \omega(782 ~\mathrm{MeV}),\quad K^{*\pm},K^{*0}, \bar{K}^{*0}(894 ~\mathrm{MeV}).
\end{align}
Their masses lie between those of the pseudoscalar mesons and baryons. These mesons can be treated as either light or heavy degrees of freedom, depending on the context. In this paper, we take $V^\mu=\mathcal{O}(p^0)$. The chiral dimensions of the other building blocks are the same as those in the pseudoscalar mesonic chiral Lagrangians \cite{leupold_towards_2012,harada_hidden_2003}. Table \ref{blbt} also provides a summary of these power-counting rules.

Additionally, each covariant derivative on the first six building blocks contributes an additional $\mathcal{O}(p^1)$ order. For the anomalous part, the Levi-Civita tensor $\varepsilon^{\mu\nu\lambda\rho}$ is also required. The above power-counting rule is effective at tree level. For loop contributions, the complex-mass scheme should be applied, as discussed in the introduction.

\subsection{Constraint relations}\label{vcr}
The chiral Lagrangian needs to satisfy all the symmetries discussed above. It is straightforward to construct a complete set that satisfies all these symmetries, but the terms in this set are generally not linearly independent. To obtain a linearly independent chiral Lagrangian, the following linear constraint relations must be taken into account \cite{fearing_extension_1996,bijnens_mesonic_1999,bijnens_anomalous_2002,bijnens_order_2018,ecker_chiral_1989,ecker_role_1989}.

\begin{enumerate}
  \item \textbf{Partial Integration.} Since derivatives acting on the entire Lagrangian can be discarded, we arrive at the following identity:
\begin{align}
\la\nabla^{\mu} A B \cdots\ra \la C \cdots\ra + \la A \nabla^{\mu} B \cdots\ra \la C \cdots\ra + \la A B \cdots\ra \la \nabla^{\mu} C \cdots\ra + \text{other terms} = 0, \label{pir}
\end{align}
where $A$, $B$, and $C$ are generic building blocks, ``$\la\cdots\ra$'' denotes the flavor trace, and ``$\cdots$'' represents additional blocks as needed. This relation implies that one of the terms in Eq.~\eqref{pir} is redundant and may be omitted. Since our focus is on single-vector-meson vertices, the covariant derivatives originally applied to $V^\mu$ can be systematically shifted to the other building blocks, eliminating the need for terms involving $\hat{V}^{\mu\nu}$ or $\nabla^\mu V^\nu$.

  \item \textbf{Equations of Motion (EOMs).} The leading-order EOMs are essential for constructing the chiral Lagrangians \cite{bijnens_mesonic_1999}. For pseudoscalar mesons, the leading-order EOM is
\begin{align}
\nabla^{\mu} u_{\mu} = \frac{i}{2} \left(\chi_{-} - \frac{1}{N_f} \left\la \chi_{-} \right\ra \right).
\end{align}
With this relation, $\nabla^{\mu} u_{\mu}$ can be replaced by other forms, and thus it does not appear in the chiral Lagrangian.

For vector mesons, the leading-order Lagrangian is given by \cite{ecker_chiral_1989}
\begin{align}
\mathcal{L}_{2} = -\frac{1}{4} \left\la \hat{V}_{\mu \nu} \hat{V}^{\mu \nu} - 2 M_V^{2} V_{\mu} V^{\mu} \right\ra.
\end{align}
This Lagrangian leads to the Klein-Gordon equation and an auxiliary condition:
\begin{align}
(\nabla_{\nu} \nabla^{\nu} + M_V^2) V^{\mu} \doteq 0, \quad \nabla_\mu V^\mu \doteq 0, \label{eomv}
\end{align}
where ``$\doteq$'' denotes equivalence up to the higher-order terms or terms of the same order with fewer covariant derivatives. This notation is meaningful because terms of the same order but with fewer derivatives are already included according to our procedure for the listing of allowable terms. These equations imply that $\nabla_{\nu} \nabla^{\nu} V^{\mu}$ and $\nabla_\mu V^\mu$ can be transformed into other forms. By applying partial integration, Eq.~\eqref{eomv} can be reformulated as
\begin{align}
V^{\mu} \nabla^{\rho} \nabla_{\rho} O_{\mu} \doteq 0, \quad V^{\mu} \nabla_{\mu} O \doteq 0,
\end{align}
where $O$ and $O_{\mu}$ denote any products of building blocks with the appropriate indices. The omitted terms do not affect the structural integrity of the chiral Lagrangian.

  \item  \textbf{Bianchi Identity.} Using the definitions in Sec.~\ref{bbp}, the following Bianchi identity is derived,
\begin{align}
\nabla^{\mu} \Gamma^{\nu \rho} + \nabla^{\nu} \Gamma^{\rho \mu} + \nabla^{\rho} \Gamma^{\mu \nu} = 0,
\end{align}
where, for any building block $X$,
\begin{align}
\left[\nabla^{\mu}, \nabla^{\nu}\right] X &= \left[\Gamma^{\mu \nu}, X\right], \\
\Gamma^{\mu \nu} &= \frac{1}{4} \left[u^{\mu}, u^{\nu}\right] - \frac{i}{2} f_{+}^{\mu \nu}.
\end{align}
Two specific cases are
\begin{align}
\nabla^{\mu} f_{-}^{\nu \alpha} + \nabla^{\nu} f_{-}^{\alpha \mu} + \nabla^{\alpha} f_{-}^{\mu \nu} &= \frac{i}{2} \left(\left[f_{+}^{\mu \nu}, u^{\alpha}\right] + \left[f_{+}^{\nu \alpha}, u^{\mu}\right] + \left[f_{+}^{\alpha \mu}, u^{\nu}\right]\right), \\
\nabla^{\mu} f_{+}^{\nu \alpha} + \nabla^{\nu} f_{+}^{\alpha \mu} + \nabla^{\alpha} f_{+}^{\mu \nu} &= \frac{i}{2} \left(\left[f_{-}^{\mu \nu}, u^{\alpha}\right] + \left[f_{-}^{\nu \alpha}, u^{\mu}\right] + \left[f_{-}^{\alpha \mu}, u^{\nu}\right]\right).
\end{align}
  \item \textbf{Schouten Identity.} In the anomalous parts, the Levi-Civita tensor $\epsilon^{\mu \nu \alpha \beta}$ is used. Since there does not exist any totally antisymmetric fifth-order tensor in the four-dimensional spacetime, the Schouten identity implies
    \begin{align}
    S^{\gamma} \epsilon^{\mu \nu \alpha \beta} - S^{\mu} \epsilon^{\gamma \nu \alpha \beta} - S^{\nu} \epsilon^{\mu \gamma \alpha \beta} - S^{\alpha} \epsilon^{\mu \nu \gamma \beta} - S^{\beta} \epsilon^{\mu \nu \alpha \gamma} = 0,
    \end{align}
    where $S^{\mu}$ denotes any operator.

    \item \textbf{Cayley-Hamilton Relations.} All building blocks are $N_f \times N_f$ matrices in the flavor space. The Cayley-Hamilton theorem provides relations among these matrices.\\
    For SU(2),
    \begin{align}
    \{A, B\} = A\la B\ra + B\la A\ra + \la A B\ra - \la A\ra\la B\ra, \label{eq.240}
    \end{align}
    where $A$ and $B$ are any $2 \times 2$ matrices.

    For SU(3),
     \begin{align}
  \begin{array}{l}{A B C+A C B+B A C+B C A+C A B+C B A-A B\la C\ra-} \\ {\quad-A C\la B\ra- B A\la C\ra- B C\la A\ra- C A\la B\ra- C B\la A\ra- A\la B C\ra-} \\ {\quad-B\la A C\ra- C\la A B\ra-\la A B C\ra-\la A C B\ra+ A\la B\ra\la C\ra+ B\la A\ra\la C\ra+} \\ {\quad+C\la A\ra\la B\ra+\la A\ra\la B C\ra+\la B\ra\la A C\ra+\la C\ra\la A B\ra-\la A\ra\la B\ra\la C\ra= 0},\end{array}\label{eq.24}
  \end{align}
    where $A$, $B$, and $C$ are any $3 \times 3$ matrices.
\end{enumerate}

\section{Tensor-field representation}
The tensor-field representation is similar to the vector-field representation. The building blocks are almost the same as those in Table \ref{blbt}, except that $V^\mu$ is replaced by $W^{\mu\nu}$. The transformation properties and power-counting rules of $W^{\mu\nu}$ are also given in Table \ref{blbt}.

In the tensor-field representation, the linear relations closely resemble those in the vector-field representation, with the primary difference being the EOMs for vector mesons. The tensor field $W^{\mu\nu}$ is antisymmetric, satisfying $W^{\mu\nu} = -W^{\nu\mu}$. Using the methods outlined in Refs. \cite{capri_second_1987, ecker_role_1989, bruns_infrared_2005}, Eq. \eqref{eomv} is replaced by
\begin{align}
(\nabla_{\rho}\nabla^{\rho} + M_{V}^2) W^{\mu\nu} \doteq 0, \quad \nabla^{\lambda} W^{\rho \sigma} + \nabla^{\rho} W^{\sigma \lambda} + \nabla^{\sigma} W^{\lambda \rho} \doteq 0. \label{eomw}
\end{align}
Additionally, there is an auxiliary condition given by \cite{capri_second_1987, rosell_quantum_2007, ecker_role_1989, bruns_infrared_2005, bijnens_tensor_1996}
\begin{align}
\nabla_{\rho} W^{\rho \mu} \doteq 0.
\end{align}
For historical reasons, we follow the convention in Eq. \eqref{eomw}, where the degrees of freedom $W^{ij}$ are fixed, while $W^{0i}$ are dynamical. By applying partial integration, Eq. \eqref{eomw} yields the following two linear relations,
\begin{align}
W_{\mu\nu} \nabla_{\rho} (R^{\mu\nu\rho} + R^{\nu\rho\mu} + R^{\rho\mu\nu}) \doteq 0, \quad
W_{\mu\nu} \nabla_{\rho} \nabla^{\rho} C^{\mu\nu} \doteq 0, \label{a.4}
\end{align}
where $R^{\mu\nu\rho}$ is any operator with three indices, and $C^{\mu\nu}$ is any operator with two indices.

\section{HLS representation}\label{ha}
The HLS representation is significantly different from the vector-field and tensor-field representations. This section provides only a brief overview, and more details can be found in Refs. \cite{bando_is_1985, bando_composite_1985, bando_vector_1985, bando_nonlinear_1988, tanabashi_chiral_1993, tanabashi_formulations_1996, harada_hidden_2003}.

\subsection{Definitions}
In the HLS representation, vector mesons are treated as gauge fields, with their masses arising from the Brout-Englert-Higgs mechanism. The HLS representation is built on the $G_{\text{global}} \times H_{\text{local}}$ symmetry, where $G = SU(N_f)_L \times SU(N_f)_R$ represents the global symmetry, and $H$ denotes the HLS. This $G_{\text{global}} \times H_{\text{local}}$ symmetry can be broken down into an $H$ symmetry, corresponding to flavor symmetry. The $U$ field from ChPT can be decomposed into two fields, $\xi_L$ and $\xi_R$, as
\begin{align}
U = \xi_L^\dagger \xi_R.
\end{align}
The two fields $\xi_L$ and $\xi_R$ satisfy the following transformation properties under the $G_{\text{global}}\times H_{\text{local}}$ symmetry,
\begin{align}
\xi_{L,R}(x)\rightarrow h(x)\cdot \xi_{L,R}(x)\cdot g_{L,R}^\dagger,\label{eq.34}
\end{align}
where $h(x)\in H_{\text{local}}$ and $g_{L,R}\in SU(N_f)_L\times SU(N_f)_R$. Two Maurer-Cartan 1-forms are
\begin{align}
\alpha_\perp^{\mu}&=\frac{1}{2i}(\partial^{\mu}\xi_R\cdot\xi_R^{\dagger}-\partial^{\mu}\xi_L\cdot\xi_L^\dagger),\\
\alpha_\parallel^{\mu}&=\frac{1}{2i}(\partial^{\mu}\xi_R\cdot\xi_R^{\dagger}+\partial^{\mu}\xi_L\cdot\xi_L^{\dagger}).
\end{align}
Their chiral transformation properties under the full symmetry are
\begin{align}
\alpha_\perp^{\mu}&\rightarrow h(x)\cdot\alpha_\perp^{\mu}\cdot h^{\dagger}(x),\\
\alpha_\parallel^{\mu}&\rightarrow h(x)\cdot\alpha_\perp^{\mu}\cdot h^{\dagger}(x)-i\partial^{\mu}h(x)\cdot h^{\dagger}(x).
\end{align}
The relations between $\alpha_\perp^{\mu}$, $\alpha_\parallel^{\mu}$, and the building blocks in the vector-field and tensor-field representations are $\alpha_\perp^{\mu} = \frac{1}{2} u^{\mu}$ and $\alpha_\parallel^\mu = i \Gamma^{\mu}$. The covariant derivatives of $\xi_{L,R}$ are given by
\begin{align}
D^{\mu} \xi_L &= \partial^{\mu} \xi_L - i K^{\mu} \xi_L + i \xi_L l^{\mu}, \\
D^{\mu} \xi_R &= \partial^{\mu} \xi_R - i K^{\mu} \xi_R + i \xi_R r^{\mu},
\end{align}
where $K^{\mu}$ are the gauge fields associated with $H_{\text{local}}$. Under a chiral transformation, $K^{\mu}$ transforms as
\begin{align}
K^{\mu} \rightarrow h(x) K^{\mu} h^{\dagger}(x) - i \partial^{\mu} h(x) h^{\dagger}(x).
\end{align}
Using $K^{\mu}$ fields, the covariant 1-forms are defined as
\begin{align}
\hat{\alpha}_{\perp}^{\mu} &= \frac{1}{2i} \left( D^{\mu} \xi_R \cdot \xi_R^{\dagger} - D^{\mu} \xi_L \cdot \xi_L^{\dagger} \right), \label{hlsb1} \\
\hat{\alpha}_{\parallel}^{\mu} &= \frac{1}{2i} \left( D^{\mu} \xi_R \cdot \xi_R^{\dagger} + D^{\mu} \xi_L \cdot \xi_L^{\dagger} \right) . \label{hlsb2}
\end{align}
The quantity $\hat{\alpha}_{\parallel}^{\mu}$ relates to the vector field $V^{\mu}$ in the standard vector-field representation as
\begin{align}
\xi \hat{\alpha}_{\parallel}^{\mu} = V^{\mu},
\end{align}
where $\xi$ is a parameter associated with the redefinition of $V^{\mu}$ in the HLS representation. The chiral transformation properties of $\hat{\alpha}_{\perp}^{\mu}$ and $\hat{\alpha}_{\parallel}^{\mu}$ are
\begin{align}
\hat{\alpha}_{\perp,\parallel}^{\mu} \rightarrow h(x) \hat{\alpha}_{\perp,\parallel}^{\mu} h^{\dagger}(x).
\end{align}
The field strength of the HLS gauge boson is defined as
\begin{align} \label{defV}
V^{\mu\nu} = \partial^{\mu} K^{\nu} - \partial^{\nu} K^{\mu} - i \left[ K^{\mu}, K^{\nu} \right].
\end{align}
The leading-order chiral Lagrangian in this representation is
\begin{align}
\mathcal{L}_{\text{H},2} = F_\pi^2 \langle \hat{\alpha}_{\perp \mu} \hat{\alpha}_{\perp}^{\mu} \rangle + F_\sigma^2 \langle \hat{\alpha}_{\parallel \mu} \hat{\alpha}_{\parallel}^{\mu} \rangle - \frac{1}{2g^2} \langle V_{\mu\nu} V^{\mu\nu} \rangle + \frac{1}{4} F_\chi^2 \langle \hat{\chi}_+ \rangle,
\end{align}
where $F_\pi$ and $F_\sigma$ are decay constants, $g$ is the HLS gauge coupling constant, and $\hat{\chi}_+$ is defined in terms of the scalar and pseudoscalar external sources as
\begin{align}
\hat{\chi}_{\pm} = \hat{\chi} \pm \hat{\chi}^{\dagger}, \quad \text{with } \hat{\chi} = \xi_L \chi \xi_R^\dagger.\label{hlsb3}
\end{align}

\subsection{Building blocks and transformation properties}

The leading-order chiral Lagrangian includes four essential building blocks: $\hat{\alpha}_\perp^{\mu}$, $\hat{\alpha}_\parallel^{\mu}$, $V^{\mu\nu}$, and $\hat{\chi}_+$. However, these building blocks alone are insufficient for constructing higher-order chiral Lagrangians. Additional terms such as $\hat{\chi}_-$ and the following building blocks are necessary:
\begin{align}
\mathcal{A}^{\mu\nu} &= D^\mu \alpha_\perp^\nu - D^\nu \alpha_\perp^\mu - i[\alpha_\parallel^\mu, \alpha_\perp^\nu] - i[\alpha_\perp^\mu, \alpha_\parallel^\nu], \\
\mathcal{H}^{\mu\nu} &= D^{\mu} \alpha_\perp^{\nu} + D^{\nu} \alpha_\perp^{\mu} - i[\alpha_\parallel^{\mu}, \alpha_\perp^{\nu}] + i[\alpha_\perp^{\mu}, \alpha_\parallel^{\nu}], \\
\mathcal{V}^{\mu\nu} &= D^\mu \alpha_\parallel^\nu - D^\nu \alpha_\parallel^\mu - i[\alpha_\parallel^\mu, \alpha_\parallel^\nu] - i[\alpha_\perp^\mu, \alpha_\perp^\nu] + V^{\mu\nu}, \\
H^{\mu\nu} &= D^\mu \alpha_\parallel^\nu + D^\nu \alpha_\parallel^\mu. \label{hlsb4}
\end{align}
These equations yield the symmetric and antisymmetric indices for $D^\mu \alpha_\perp^\nu$ and $D^\mu \alpha_\parallel^\nu$. At NLO, only the antisymmetric building blocks $\mathcal{A}^{\mu\nu}$ and $\mathcal{V}^{\mu\nu}$ are present, as noted in Ref.~\cite{harada_hidden_2003}. At higher orders, the symmetric forms of $D^\mu \alpha_\perp^\nu$ and $D^\mu \alpha_\parallel^\nu$ are also needed. This paper denotes these symmetric tensors as $\mathcal{H}^{\mu\nu}$ and $H^{\mu\nu}$, respectively. Additionally, an alternative set of building blocks has been proposed in the literature~\cite{ma_hidden_2005}, but we adopt the former convention for consistency. Under a chiral rotation, any building block $Y$ transforms as $Y \rightarrow h Y h^\dagger$. The covariant derivative of these building blocks is given by
\begin{align}
D^\mu Y = \partial^\mu Y - i[K^\mu, Y].
\end{align}

The structure of these building blocks in the HLS representation differs significantly from those in the other representations. In general, the relations among these building blocks are quite complex. However, the situation can be simplified in unitary gauge, as discussed in Refs.~\cite{tanabashi_chiral_1993, tanabashi_formulations_1996, harada_hidden_2003}. Table~\ref{blhbt} outlines the relations between Eqs.~\eqref{hlsb1}, \eqref{hlsb2}, (\ref{hlsb3}-\ref{hlsb4}), and Eq.~\eqref{bb} in the unitary gauge, along with the transformation properties of the building blocks. With these building blocks and their covariant derivatives, the chiral Lagrangians at any order can be constructed.
\begin{table*}[!h]
\caption{\label{blhbt}Chiral dimension (Dim), relations with pseudoscalar mesonic building blocks $O$, parity ($P$), charge conjugation ($C$) and hermiticity (h.c.) of the building blocks in the HLS representation.}
\begin{center}
\begin{tabular}{cccccc}
\hline\hline
& Dim &                               O               &  $P$               &                $C$                &               h.c.                \\
\hline
$\ahe^{\mu}$            &  1  &                      $\frac{1}{2}u^{\mu}$   &   $-\ahe_{\mu}$                   &         $(\ahe^{\mu})^T$          &           $\ahe^{\mu}$            \\
$\ahp^{\mu}$            &  1  &                  &  $\ahp_{\mu}$              &         $-(\ahp^{\mu})^T$         &           $\ahp^{\mu}$            \\
$\hat\chi_{\pm}$        &  2  &                          $\chi_{\pm}$         &   $\hat\chi_{\pm} $             &       $(\hat\chi_{\pm})^T$        &       $\pm \hat\chi_{\pm}$        \\
$\Vs^{\mu\nu}$           &  2  &                   $\frac{1}{2}\fp^{\mu\nu}$      &   $\Vs_{\mu\nu}$             &        $-(\Vs^{\mu\nu})^T$        &          $ \Vs^{\mu\nu}$          \\
$\Vt^{\mu\nu}$           &  2  &    &$\Vt_{\mu\nu}$  &        $-(\Vt^{\mu\nu})^T$        &          $\Vt^{\mu\nu}$           \\
$\Hs^{\mu\nu}$           &  2  &                    $\frac{1}{2}h^{\mu\nu}$      &$-\Hs_{\mu\nu}$               &        $(\Hs^{\mu\nu})^T$         &          $\Hs^{\mu\nu}$           \\
$\As^{\mu\nu}$           &  2  &                   $-\frac{1}{2}\fm^{\mu\nu}$     &$-\As_{\mu\nu}$               &        $(\As^{\mu\nu})^T$         &          $\As^{\mu\nu}$           \\
$H^{\mu\nu}$            &  2  &                                                            &$H_{\mu\nu}$      &          $-(H^{\mu\nu})^T$         &           $H^{\mu\nu}$            \\
$\varepsilon^{\mu\nu\lambda\rho}$ &  0  &                                                  &  $-\varepsilon_{\mu\nu\lambda\rho}$             & $\varepsilon^{\mu\nu\lambda\rho}$ & $\varepsilon^{\mu\nu\lambda\rho}$ \\
\hline\hline
\end{tabular}
\end{center}
\end{table*}

Table~\ref{blhbt} also provides the power counting of the HLS building blocks, as described in Refs.~\cite{tanabashi_chiral_1993, harada_hidden_2003}. Furthermore, the parameter $g$ in the leading-order chiral Lagrangian is assigned a chiral order of $\mathcal{O}(p^1)$~\cite{harada_hidden_2003}.

\subsection{Constraint relations}

For convenience, we discuss all constraint relations in the unitary gauge. As a result, most conclusions from Sec.~\ref{vcr} remain applicable, provided that the relations in Table~\ref{blhbt} are employed. One notable difference lies in the EOMs, as detailed in Ref.~\cite{harada_hidden_2003}. The EOMs in the HLS representation are given by
\begin{align}
&D_\mu\ahe^\mu=-i(a-1)[\ahp_\mu,\ahe^\mu]-\frac{i}{4}\frac{F_\chi^2}{F_\pi^2}\bigg(\chimh-\frac{1}{N_f}\la\chimh\ra\bigg)+\mathcal{O}(p^4),\notag\\
&D_\mu\ahp^\mu=\mathcal{O}(p^4),\notag\\
&D_{\nu}V^{\nu\mu}=g^2 F_\sigma^2\ahp^\mu+\mathcal{O}(p^4).\label{eq.23}
\end{align}
Here, the parameter $a$ is defined as $a = F_\sigma^2 / F_\pi^2$.

For $V^{\mu\nu}$, with the definition provided in Eq.~\eqref{defV}, a relation analogous to the Bianchi identity can be derived:
\begin{align}
&D^\rho V^{\mu\nu} + D^\mu V^{\nu\rho} + D^\nu V^{\rho\mu} = 0. \label{eq.28}
\end{align}
Thus, we have established all linear relations within the HLS representation.

\section{Results of chiral Lagrangians with vector mesons}\label{rcl}
The leading-order (LO) chiral Lagrangians incorporating vector meson octet were derived over 30 years ago \cite{ecker_chiral_1989, ecker_role_1989, bando_is_1985}. In this work, we extend the discussion to nonet and the higher-order Lagrangians. However, these higher-order Lagrangians involve a vast number of terms, and manually performing the calculations can lead to the inadvertent omission of some linear relations. Consequently, all calculations are performed with the aid of a computer.

One major challenge in these computations is symbolic manipulation. So far, to our knowledge, no effective and widely accessible program exists to automatically generate all possible terms and associated linear relations. To overcome this, we employ a trick that converts symbolic computations into numerical ones. This paper outlines the basic approach, with detailed explanations provided in Refs.~\cite{Jiang:2014via, Jiang:2016vax, Jiang:2017yda}.

To avoid symbolic complications, we assign unique numbers to each building block and index. For instance, we may assign $V \to 11$, $u \to 12$, $\mu \to 1$, $\nu \to 2$, and so on. Using this scheme, a term like $V^{\mu} u_{\mu} u^{\nu} u_{\nu}$ can be represented by two vectors: one for the building blocks, $(11, 12, 12, 12)$, and another for the indices, $(1, 1, 2, 2)$. Compared to handling symbolic expressions, processing these vectors computationally is significantly more manageable. As a result, most of the tedious calculations are efficiently executed using this numerical approach.

\subsection{LO}
The LO chiral Lagrangians are simple. We only give a simple summary and add the singlet state. For the vector-field representation, the independent terms are given in Table \ref{vector LO}. The $\la V_{\mu}\ra$ terms, such as the fourth term in the $SU(2)$ case, only contain the singlet state.

\begin{table*}[!h]
\caption{\label{vector LO}The LO results for the vector-field representation. The numbers are the sequence numbers in each case.}
\begin{center}
\begin{tabular}{lcc}
\hline\hline terms & $SU(2)$ & $SU(3)$\\
\hline
$i\la V^{\mu}u^{\nu}f_{-\mu\nu}\ra+\mathrm{H.c.}$ & 1 & 1  \\
$i\la V^{\mu}u^{\nu}h_{\mu\nu}\ra+\mathrm{H.c.}$ & 2 & 2  \\
$\la V^{\mu}\nabla^{\nu}f_{+\mu\nu}\ra$ & 3 & 3  \\
$\la V^{\mu}\ra\la\nabla^{\nu}f_{+\mu\nu}\ra$ & 4 &  \\
$\la V^{\mu}u_{\mu}\chim\ra+\mathrm{H.c.}$ & 5 & 4  \\
$i\varepsilon^{\mu\nu\lambda\rho}\la V_{\mu}u_{\nu}u_{\lambda}u_{\rho}\ra$ & 6 & 5  \\
$i\varepsilon^{\mu\nu\lambda\rho}\la V_{\mu}\ra\la u_{\nu}u_{\lambda}u_{\rho}\ra$ &  & 6  \\
$\varepsilon^{\mu\nu\lambda\rho}\la V_{\mu}f_{+\nu\lambda}u_{\rho}\ra+\mathrm{H.c.}$ & 7 & 7  \\
$\varepsilon^{\mu\nu\lambda\rho}\la V_{\mu}\ra\la f_{+\nu\lambda}u_{\rho}\ra$ & 8 & 8  \\
\hline\hline
\end{tabular}
\end{center}
\end{table*}

For the tensor-field representation, the independent terms are given in Table \ref{tensor LO}. The third term in the $SU(2)$ case only contains the singlet state.

\begin{table*}[!h]
\caption{\label{tensor LO}The LO results for the tensor-field representation. The numbers are the sequence numbers in each case.}
\begin{center}
\begin{tabular}{lcc}
\hline\hline terms & $SU(2)$ & $SU(3)$\\
\hline
$i\la W^{\mu\nu}u_{\mu}u_{\nu}\ra$ & 1 & 1  \\
$\la W^{\mu\nu}f_{+\mu\nu}\ra$ & 2 & 2  \\
$\la W^{\mu\nu}\ra\la f_{+\mu\nu}\ra$ & 3 &  \\
\hline\hline
\end{tabular}
\end{center}
\end{table*}

For the HLS representation, the independent terms are given in Table \ref{HLS LO}. The third term in both cases only contains the singlet state.

\begin{table*}[!h]
\caption{\label{HLS LO}The LO results for the HLS representation. The numbers are the sequence numbers in each case.}
\begin{center}
\begin{tabular}{lcc}
\hline\hline terms & $SU(2)$ & $SU(3)$\\
\hline
$\la{\hat{a}_{\perp}}^{\mu}\hat{a}_{\perp\mu}\ra$ & 1 & 1  \\
$\la{\hat{a}_{\parallel}}^{\mu}\hat{a}_{\parallel\mu}\ra$ & 2 & 2  \\
$\la{\hat{a}_{\parallel}}^{\mu}\ra\la\hat{a}_{\parallel\mu}\ra$ & 3 & 3 \\
$\la\chiph\ra$ & 4 & 4  \\
\hline\hline
\end{tabular}
\end{center}
\end{table*}

\subsection{NLO}

Because the expressions are very lengthy, the NLO result for the vector-field representation is provided in Tables \ref{vnlo} in the supplementary material. The tensor-field and HLS representations are given in Tables \ref{tnlo} and \ref{hnlo}, respectively.

\begin{table}[!h]
\caption{The NLO chiral Lagrangians with vector mesons in the tensor-field representation.}\label{tnlo}
\begin{tabular}{lcclcc}
\hline\hline $P_n$ & $SU(2)$ & $SU(3)$ & $P_n$ & $SU(2)$ & $SU(3)$\\
\hline
$i\la W^{\mu\nu}u_{\mu}u_{\nu}u^{\lambda}u_{\lambda}\ra+\mathrm{H.c.}$ & 1 & 1 & $\la{f_{+}}^{\mu\nu}\ra\la W_{\mu\nu}\chip\ra$ & 19 &  \\
$i\la W^{\mu\nu}u_{\mu}u^{\lambda}u_{\nu}u_{\lambda}\ra+\mathrm{H.c.}$ & 2 & 2 & $\la W^{\mu\nu}f_{-\mu\nu}\chim\ra+\mathrm{H.c.}$ & 20 & 28  \\
$i\la W^{\mu\nu}u_{\mu}u^{\lambda}u_{\lambda}u_{\nu}\ra$ &  & 3 & $\la W^{\mu\nu}u_{\mu}\nabla_{\nu}\chi_{-}\ra+\mathrm{H.c.}$ & 21 & 29  \\
$i\la W^{\mu\nu}u^{\lambda}u_{\mu}u_{\nu}u_{\lambda}\ra$ &  & 4 & $i\varepsilon^{\mu\nu\lambda\rho}\la W_{\mu\nu}u_{\lambda}u^{\sigma}f_{-\rho\sigma}\ra+\mathrm{H.c.}$ & 22 & 30  \\
$i\la W^{\mu\nu}\ra\la u_{\mu}u_{\nu}u^{\lambda}u_{\lambda}\ra$ &  & 5 & $i\varepsilon^{\mu\nu\lambda\rho}\la W_{\mu\nu}u^{\sigma}u_{\lambda}f_{-\rho\sigma}\ra+\mathrm{H.c.}$ &  & 31  \\
$i\la u^{\mu}u_{\mu}\ra\la W^{\nu\lambda}u_{\nu}u_{\lambda}\ra$ &  & 6 & $i\varepsilon^{\mu\nu\lambda\rho}\la W_{\mu\nu}u^{\sigma}u_{\sigma}f_{-\lambda\rho}\ra+\mathrm{H.c.}$ &  & 32  \\
$i\la W^{\mu\nu}{f_{-\mu}}^{\lambda}f_{-\nu\lambda}\ra$ & 3 & 7 & $i\varepsilon^{\mu\nu\lambda\rho}\la W_{\mu\nu}u_{\lambda}{f_{-\rho}}^{\sigma}u_{\sigma}\ra+\mathrm{H.c.}$ &  & 33  \\
$i\la W^{\mu\nu}{f_{-\mu}}^{\lambda}h_{\nu\lambda}\ra+\mathrm{H.c.}$ & 4 & 8 & $i\varepsilon^{\mu\nu\lambda\rho}\la W_{\mu\nu}u_{\lambda}u^{\sigma}h_{\rho\sigma}\ra+\mathrm{H.c.}$ & 23 & 34  \\
$i\la W^{\mu\nu}{h_{\mu}}^{\lambda}h_{\nu\lambda}\ra$ & 5 & 9 & $i\varepsilon^{\mu\nu\lambda\rho}\la W_{\mu\nu}u^{\sigma}u_{\lambda}h_{\rho\sigma}\ra+\mathrm{H.c.}$ &  & 35  \\
$\la W^{\mu\nu}f_{+\mu\nu}u^{\lambda}u_{\lambda}\ra+\mathrm{H.c.}$ & 6 & 10 & $i\varepsilon^{\mu\nu\lambda\rho}\la W_{\mu\nu}\ra\la u_{\lambda}u^{\sigma}f_{-\rho\sigma}\ra+\mathrm{H.c.}$ &  & 36  \\
$\la W^{\mu\nu}{f_{+\mu}}^{\lambda}u_{\nu}u_{\lambda}\ra+\mathrm{H.c.}$ & 7 & 11 & $i\varepsilon^{\mu\nu\lambda\rho}\la W_{\mu\nu}\ra\la u_{\lambda}u^{\sigma}h_{\rho\sigma}\ra+\mathrm{H.c.}$ &  & 37  \\
$\la W^{\mu\nu}{f_{+\mu}}^{\lambda}u_{\lambda}u_{\nu}\ra+\mathrm{H.c.}$ & 8 & 12 & $\varepsilon^{\mu\nu\lambda\rho}\la W_{\mu\nu}{f_{+\lambda}}^{\sigma}f_{-\rho\sigma}\ra+\mathrm{H.c.}$ & 24 & 38  \\
$\la W^{\mu\nu}u_{\mu}{f_{+\nu}}^{\lambda}u_{\lambda}\ra+\mathrm{H.c.}$ & 9 & 13 & $\varepsilon^{\mu\nu\lambda\rho}\la W_{\mu\nu}{f_{+\lambda}}^{\sigma}h_{\rho\sigma}\ra+\mathrm{H.c.}$ & 25 & 39  \\
$\la W^{\mu\nu}u^{\lambda}f_{+\mu\nu}u_{\lambda}\ra$ & 10 & 14 & $\varepsilon^{\mu\nu\lambda\rho}\la W_{\mu\nu}\nabla^{\sigma}f_{+\lambda\sigma}u_{\rho}\ra+\mathrm{H.c.}$ & 26 & 40  \\
$\la W^{\mu\nu}\ra\la f_{+\mu\nu}u^{\lambda}u_{\lambda}\ra$ & 11 & 15 & $\varepsilon^{\mu\nu\lambda\rho}\la W_{\mu\nu}\ra\la{f_{+\lambda}}^{\sigma}f_{-\rho\sigma}\ra$ & 27 & 41  \\
$\la W^{\mu\nu}f_{+\mu\nu}\ra\la u^{\lambda}u_{\lambda}\ra$ &  & 16 & $\varepsilon^{\mu\nu\lambda\rho}\la W_{\mu\nu}\ra\la{f_{+\lambda}}^{\sigma}h_{\rho\sigma}\ra$ & 28 & 42  \\
$\la W^{\mu\nu}\ra\la{f_{+\mu}}^{\lambda}u_{\nu}u_{\lambda}\ra+\mathrm{H.c.}$ & 12 & 17 & $\varepsilon^{\mu\nu\lambda\rho}\la W_{\mu\nu}\ra\la\nabla^{\sigma}f_{+\lambda\sigma}u_{\rho}\ra$ & 29 & 43  \\
$\la W^{\mu\nu}{f_{+\mu}}^{\lambda}\ra\la u_{\nu}u_{\lambda}\ra$ &  & 18 & $i\varepsilon^{\mu\nu\lambda\rho}\la W_{\mu\nu}f_{-\lambda\rho}\chip\ra+\mathrm{H.c.}$ & 30 & 44  \\
$\la W^{\mu\nu}u_{\mu}\ra\la{f_{+\nu}}^{\lambda}u_{\lambda}\ra$ &  & 19 & $\varepsilon^{\mu\nu\lambda\rho}\la W_{\mu\nu}u_{\lambda}u_{\rho}\chim\ra+\mathrm{H.c.}$ & 31 & 45  \\
$i\la W^{\mu\nu}{f_{+\mu}}^{\lambda}f_{+\nu\lambda}\ra$ & 13 & 20 & $\varepsilon^{\mu\nu\lambda\rho}\la W_{\mu\nu}u_{\lambda}\chim u_{\rho}\ra$ & 32 & 46  \\
$i\la W^{\mu\nu}u_{\mu}u_{\nu}\chip\ra+\mathrm{H.c.}$ & 14 & 21 & $\varepsilon^{\mu\nu\lambda\rho}\la W_{\mu\nu}\ra\la u_{\lambda}u_{\rho}\chim\ra$ &  & 47  \\
$i\la W^{\mu\nu}u_{\mu}\chip u_{\nu}\ra$ & 15 & 22 & $\varepsilon^{\mu\nu\lambda\rho}\la\chim\ra\la W_{\mu\nu}u_{\lambda}u_{\rho}\ra$ &  & 48  \\
$i\la W^{\mu\nu}\ra\la u_{\mu}u_{\nu}\chip\ra$ &  & 23 & $i\varepsilon^{\mu\nu\lambda\rho}\la W_{\mu\nu}f_{+\lambda\rho}\chim\ra+\mathrm{H.c.}$ & 33 & 49  \\
$i\la\chip\ra\la W^{\mu\nu}u_{\mu}u_{\nu}\ra$ &  & 24 & $i\varepsilon^{\mu\nu\lambda\rho}\la W_{\mu\nu}\ra\la f_{+\lambda\rho}\chim\ra$ & 34 & 50  \\
$\la W^{\mu\nu}f_{+\mu\nu}\chip\ra+\mathrm{H.c.}$ & 16 & 25 & $i\varepsilon^{\mu\nu\lambda\rho}\la\chim\ra\la W_{\mu\nu}f_{+\lambda\rho}\ra$ & 35 & 51  \\
$\la W^{\mu\nu}\ra\la f_{+\mu\nu}\chip\ra$ & 17 & 26 & $i\varepsilon^{\mu\nu\lambda\rho}\la f_{+\mu\nu}\ra\la W_{\lambda\rho}\chim\ra$ & 36 &  \\
$\la\chip\ra\la W^{\mu\nu}f_{+\mu\nu}\ra$ & 18 & 27  \\
\hline
\end{tabular}
\end{table}

\begin{table}[!h]
\caption{The NLO chiral Lagrangians with vector mesons in the HLS representation.}\label{hnlo}
\begin{tabular}{lcclcc}
\hline\hline $P_n$ & $SU(2)$ & $SU(3)$ & $P_n$ & $SU(2)$ & $SU(3)$\\
\hline
$\la{\hat{a}_{\perp}}^{\mu}\hat{a}_{\perp\mu}{\hat{a}_{\perp}}^{\nu}\hat{a}_{\perp\nu}\ra$ & 1 & 1 & $\la{\hat{a}_{\perp}}^{\mu}\hat{a}_{\parallel\mu}\ra\la{\hat{a}_{\perp}}^{\nu}\hat{a}_{\parallel\nu}\ra$ &  & 29  \\
$\la{\hat{a}_{\perp}}^{\mu}{\hat{a}_{\perp}}^{\nu}\hat{a}_{\perp\mu}\hat{a}_{\perp\nu}\ra$ & 2 & 2 & $\la{\hat{a}_{\perp}}^{\mu}{\hat{a}_{\parallel}}^{\nu}\ra\la\hat{a}_{\perp\mu}\hat{a}_{\parallel\nu}\ra$ &  & 30  \\
$\la{\hat{a}_{\perp}}^{\mu}\hat{a}_{\perp\mu}\ra\la{\hat{a}_{\perp}}^{\nu}\hat{a}_{\perp\nu}\ra$ &  & 3 & $\la{\hat{a}_{\perp}}^{\mu}{\hat{a}_{\parallel}}^{\nu}\ra\la\hat{a}_{\perp\nu}\hat{a}_{\parallel\mu}\ra$ &  & 31  \\
$i\la{\hat{a}_{\perp}}^{\mu}{\hat{a}_{\perp}}^{\nu}\Vs_{\mu\nu}\ra$ & 3 & 4 & $i\la{\hat{a}_{\parallel}}^{\mu}{\hat{a}_{\parallel}}^{\nu}\Vs_{\mu\nu}\ra$ & 25 & 32  \\
$\la\Vs^{\mu\nu}\Vs_{\mu\nu}\ra$ & 4 & 5 & $\la{\hat{a}_{\parallel}}^{\mu}\hat{a}_{\parallel\mu}\chiph\ra$ & 26 & 33  \\
$\la\Vs^{\mu\nu}\ra\la\Vs_{\mu\nu}\ra$ & 5 & 6 & $\la{\hat{a}_{\parallel}}^{\mu}\ra\la\hat{a}_{\parallel\mu}\chiph\ra$ & 27 & 34  \\
$\la{\hat{a}_{\perp}}^{\mu}\hat{a}_{\perp\mu}\chiph\ra$ & 6 & 7 & $\la\chiph\ra\la{\hat{a}_{\parallel}}^{\mu}\hat{a}_{\parallel\mu}\ra$ & 28 & 35  \\
$\la\chiph\ra\la{\hat{a}_{\perp}}^{\mu}\hat{a}_{\perp\mu}\ra$ &  & 8 & $\la{\hat{a}_{\parallel}}^{\mu}\ra\la\hat{a}_{\parallel\mu}\ra\la\chiph\ra$ &  & 36  \\
$\la\chiph^{2}\ra$ & 7 & 9 & $i\la{\hat{a}_{\parallel}}^{\mu}{\hat{a}_{\parallel}}^{\nu}V_{\mu\nu}\ra$ & 29 & 37  \\
$\la\chiph\ra\la\chiph\ra$ & 8 & 10 & $\la{\hat{a}_{\parallel}}^{\mu}\hat{a}_{\parallel\mu}{\hat{a}_{\parallel}}^{\nu}\hat{a}_{\parallel\nu}\ra$ & 30 & 38  \\
$\la\chimh^{2}\ra$ & 9 & 11 & $\la{\hat{a}_{\parallel}}^{\mu}{\hat{a}_{\parallel}}^{\nu}\hat{a}_{\parallel\mu}\hat{a}_{\parallel\nu}\ra$ & 31 & 39  \\
$\la\chimh\ra\la\chimh\ra$ & 10 & 12 & $\la{\hat{a}_{\parallel}}^{\mu}\ra\la\hat{a}_{\parallel\mu}{\hat{a}_{\parallel}}^{\nu}\hat{a}_{\parallel\nu}\ra$ & 32 & 40  \\
$\la\Hs^{\mu\nu}\Hs_{\mu\nu}\ra$ & 11 & 13 & $\la{\hat{a}_{\parallel}}^{\mu}\hat{a}_{\parallel\mu}\ra\la{\hat{a}_{\parallel}}^{\nu}\hat{a}_{\parallel\nu}\ra$ & 33 & 41  \\
$i\la{\hat{a}_{\perp}}^{\mu}{\hat{a}_{\perp}}^{\nu}V_{\mu\nu}\ra$ & 12 & 14 & $\la{\hat{a}_{\parallel}}^{\mu}{\hat{a}_{\parallel}}^{\nu}\ra\la\hat{a}_{\parallel\mu}\hat{a}_{\parallel\nu}\ra$ & 34 & 42  \\
$\la V^{\mu\nu}\Vs_{\mu\nu}\ra$ & 13 & 15 & $\la{\hat{a}_{\parallel}}^{\mu}\ra\la\hat{a}_{\parallel\mu}\ra\la{\hat{a}_{\parallel}}^{\nu}\hat{a}_{\parallel\nu}\ra$ &  & 43  \\
$\la V^{\mu\nu}V_{\mu\nu}\ra$ & 14 & 16 & $\la{\hat{a}_{\parallel}}^{\mu}\ra\la{\hat{a}_{\parallel}}^{\nu}\ra\la\hat{a}_{\parallel\mu}\hat{a}_{\parallel\nu}\ra$ &  & 44  \\
$\la{\hat{a}_{\perp}}^{\mu}\hat{a}_{\parallel\mu}\chimh\ra+\mathrm{H.c.}$ & 15 & 17 & $\la{\hat{a}_{\parallel}}^{\mu}\ra\la D^{\nu}H_{\mu\nu}\ra$ & 35 & 45  \\
$i\la{\hat{a}_{\perp}}^{\mu}{\hat{a}_{\parallel}}^{\nu}\Hs_{\mu\nu}\ra+\mathrm{H.c.}$ & 16 & 18 & $i\varepsilon^{\mu\nu\lambda\rho}\la\hat{a}_{\perp\mu}\hat{a}_{\perp\nu}\hat{a}_{\perp\lambda}\hat{a}_{\parallel\rho}\ra$ & 36 & 46  \\
$\la{\hat{a}_{\parallel}}^{\mu}\ra\la D^{\nu}\Vs_{\mu\nu}\ra$ & 17 & 19 & $i\varepsilon^{\mu\nu\lambda\rho}\la\hat{a}_{\parallel\mu}\ra\la\hat{a}_{\perp\nu}\hat{a}_{\perp\lambda}\hat{a}_{\perp\rho}\ra$ &  & 47  \\
$\la{\hat{a}_{\perp}}^{\mu}\hat{a}_{\perp\mu}{\hat{a}_{\parallel}}^{\nu}\hat{a}_{\parallel\nu}\ra$ & 18 & 20 & $\varepsilon^{\mu\nu\lambda\rho}\la\hat{a}_{\perp\mu}\hat{a}_{\parallel\nu}\Vs_{\lambda\rho}\ra+\mathrm{H.c.}$ & 37 & 48  \\
$\la{\hat{a}_{\perp}}^{\mu}{\hat{a}_{\perp}}^{\nu}\hat{a}_{\parallel\mu}\hat{a}_{\parallel\nu}\ra$ & 19 & 21 & $\varepsilon^{\mu\nu\lambda\rho}\la\Vs_{\mu\nu}\ra\la\hat{a}_{\perp\lambda}\hat{a}_{\parallel\rho}\ra$ & 38 & 49  \\
$\la{\hat{a}_{\perp}}^{\mu}{\hat{a}_{\perp}}^{\nu}\hat{a}_{\parallel\nu}\hat{a}_{\parallel\mu}\ra$ & 20 & 22 & $\varepsilon^{\mu\nu\lambda\rho}\la\hat{a}_{\parallel\mu}\ra\la\hat{a}_{\perp\nu}\Vs_{\lambda\rho}\ra$ &  & 50  \\
$\la{\hat{a}_{\perp}}^{\mu}\hat{a}_{\parallel\mu}{\hat{a}_{\perp}}^{\nu}\hat{a}_{\parallel\nu}\ra+\mathrm{H.c.}$ & 21 & 23 & $\varepsilon^{\mu\nu\lambda\rho}\la\hat{a}_{\perp\mu}\hat{a}_{\parallel\nu}V_{\lambda\rho}\ra+\mathrm{H.c.}$ & 39 & 51  \\
$\la{\hat{a}_{\perp}}^{\mu}{\hat{a}_{\parallel}}^{\nu}\hat{a}_{\perp\mu}\hat{a}_{\parallel\nu}\ra$ & 22 & 24 & $\varepsilon^{\mu\nu\lambda\rho}\la\hat{a}_{\parallel\mu}\ra\la\hat{a}_{\perp\nu}V_{\lambda\rho}\ra$ & 40 & 52  \\
$\la{\hat{a}_{\perp}}^{\mu}\hat{a}_{\perp\mu}\ra\la{\hat{a}_{\parallel}}^{\nu}\hat{a}_{\parallel\nu}\ra$ &  & 25 & $\varepsilon^{\mu\nu\lambda\rho}\la\hat{a}_{\parallel\mu}\ra\la\hat{a}_{\parallel\nu}\As_{\lambda\rho}\ra$ &  & 53  \\
$\la{\hat{a}_{\parallel}}^{\mu}\ra\la{\hat{a}_{\perp}}^{\nu}\hat{a}_{\perp\nu}\hat{a}_{\parallel\mu}\ra$ & 23 & 26 & $i\varepsilon^{\mu\nu\lambda\rho}\la\hat{a}_{\perp\mu}\hat{a}_{\parallel\nu}\hat{a}_{\parallel\lambda}\hat{a}_{\parallel\rho}\ra$ & 41 & 54  \\
$\la{\hat{a}_{\perp}}^{\mu}{\hat{a}_{\perp}}^{\nu}\ra\la\hat{a}_{\parallel\mu}\hat{a}_{\parallel\nu}\ra$ &  & 27 & $i\varepsilon^{\mu\nu\lambda\rho}\la\hat{a}_{\parallel\mu}\ra\la\hat{a}_{\perp\nu}\hat{a}_{\parallel\lambda}\hat{a}_{\parallel\rho}\ra$ &  & 55  \\
$\la{\hat{a}_{\parallel}}^{\mu}\ra\la{\hat{a}_{\perp}}^{\nu}\hat{a}_{\perp\mu}\hat{a}_{\parallel\nu}\ra+\mathrm{H.c.}$ & 24 & 28  \\
\hline
\end{tabular}
\end{table}

The numbers of these chiral Lagrangians are listed in Table~\ref{vector numnn}. Regarding the applications of these higher-order Lagrangian terms, in principle, one can integrate out the vector field to relate the low-energy constants in the vector meson Lagrangian to those in the pseudoscalar meson Lagrangian. A related discussion of the vector meson octet can be found in Ref. \cite{Guo:2020cmv}.

\begin{table*}[!h]
\caption{\label{vector numnn}The numbers of the NLO chiral Lagrangians with vector mesons in the different representations.}
\begin{center}
\begin{tabular}{ccccccc}
\hline\hline
      Flavor     & \multicolumn{2}{c}{$SU(2)$} & \multicolumn{3}{c}{$SU(3)$} &     \\
\hline
  Representation & Vector &       Tensor       & HLS & Vector &    Tensor    & HLS \\
      Normal     &   81   &         21         & 35  &  136   &      29      & 45  \\
     Anomaly     &   42   &         15         &  6  &   95   &      22      & 10  \\
      Total      &  123   &         36         & 41  &  231   &      51      & 55  \\
\hline\hline
\end{tabular}
\end{center}
\end{table*}

To check our results, we have compared them with existing octet results and found some redundant terms.
The results for the octet tensor-field representation in the large-$N_C$ limit are presented in Ref. \cite{cirigliano_towards_2006}. In the three-flavor case, using the linear relations from Sec. \ref{vcr}, we find an additional relation:
\begin{align}
{\cal O}^V_{17} \doteq \frac{1}{2}{\cal O}^V_{11} - {\cal O}^V_{12} - {\cal O}^V_{13} - {\cal O}^V_{15} - 2{\cal O}^V_{16}, \label{tnlo1}
\end{align}
where ${\cal O}^V_i$ denotes the $i$-th term listed in Table 1 of Ref. \cite{cirigliano_towards_2006}. Together with the relations given in Eq. \eqref{a.4}, three additional relations arise:
\begin{align}
&{\cal O}^V_{19} \doteq -2{\cal O}^V_{5} + \frac{1}{2}{\cal O}^V_{11} - {\cal O}^V_{12} - {\cal O}^V_{13} - {\cal O}^V_{15} - 2{\cal O}^V_{16} + 2{\cal O}^V_{18} - \frac{1}{2} {\cal O}^V_{20},\quad
{\cal O}^V_{21} \doteq 0,\quad {\cal O}^V_{22} \doteq 0. \label{tnlo2}
\end{align}
The terms on the left-hand side of Eqs. \eqref{tnlo1} and \eqref{tnlo2} are not independent and can be eliminated. Since the large-$N_c$ limit is not considered here, the following terms at the $\mathcal{O}(1/N_C)$ level need to be added to Ref. \cite{cirigliano_towards_2006}:
\begin{align}
&{\cal O}^V_{23} = i\la u^{\mu}u_{\mu}\ra\la W^{\nu\lambda}u_{\nu}u_{\lambda}\ra, \quad
{\cal O}^V_{24} = \la W^{\mu\nu}f_{+\mu\nu}\ra\la u^{\lambda}u_{\lambda}\ra, \quad
{\cal O}^V_{25} = \la W^{\mu\nu}{f_{+\mu}}^{\lambda}\ra\la u_{\nu}u_{\lambda}\ra, \notag\\
&{\cal O}^V_{26} = \la W^{\mu\nu}u_{\mu}\ra\la {f_{+\nu}}^{\lambda}u_{\lambda}\ra, \quad
{\cal O}^V_{27} = i\la\chip\ra\la W^{\mu\nu}u_{\mu}u_{\nu}\ra, \quad
{\cal O}^V_{28} = \la\chip\ra\la W^{\mu\nu}f_{+\mu\nu}\ra.
\end{align}
Now, only 24 independent terms are left, i.e. ${\cal O}^V_{i}, i=1,2,\cdots,28$, but $i\neq 17, 19, 21, 22$.

In the two-flavor case, only 16 terms remain, requiring the extra removal of ${\cal O}^V_{3}$, ${\cal O}^V_{4}$, ${\cal O}^V_{9}$, ${\cal O}^V_{23}$, ${\cal O}^V_{24}$, ${\cal O}^V_{25}$, ${\cal O}^V_{26}$, and ${\cal O}^V_{27}$.

For the anomalous octet part, the complete results have been provided in Refs. \cite{Kampf:2011ty,Roig:2013baa}. Using the linear relations from Sec. \ref{vcr} and Eq. \eqref{a.4}, three additional relations are identified:
\begin{align}
{\cal O}^{V,a}_{2} \doteq -{\cal O}^{V,a}_{1} - {\cal O}^{V,a}_{5} + {\cal O}^{V,a}_{7} + {\cal O}^{V,a}_{8}, \quad {\cal O}^{V,a}_{4} \doteq -\frac{1}{2} {\cal O}^{V,a}_{15}, \quad {\cal O}^{V,a}_{16} \doteq -\frac{1}{2} {\cal O}^{V,a}_{11} + \frac{1}{2} {\cal O}^{V,a}_{12}.
\end{align}
where ${\cal O}^{V,a}_i$ denotes the $i$-th term in Eq. (25) of Ref. \cite{Kampf:2011ty}.

In the HLS representation for octet, we reproduce the results from Refs. \cite{tanabashi_chiral_1993,harada_hidden_2003}, with a necessary correction. Specifically, for the three-flavor normal terms, the $y_{17}$ term from Ref. \cite{harada_hidden_2003} is related to other terms through the Cayley-Hamilton relation \eqref{eq.24}:
\begin{align}
Y_{17} = 2Y_6 + 2Y_7 + Y_8 + Y_{15} + Y_{16},
\end{align}
where $Y_i$ corresponds to the term associated with $y_i$. This relation arises because Eq. \eqref{eq.24} is more general. The form used in Refs. \cite{tanabashi_chiral_1993,harada_hidden_2003} is a special case of Eq. \eqref{eq.24}. Further details are provided in Sec. 4.3 of Ref. \cite{harada_hidden_2003}.

For the three-flavor anomalous terms, some literature has given similar results with different building blocks \cite{fujiwara_non-abelian_1985,Furui:1986ep,Jain:1987sz,bando_nonlinear_1988,harada_hidden_2003}. With these different building blocks, the results are
\begin{align}
\mathcal{L}_{\text{H},A,4}=&\frac{N_c}{16\pi^2}\big[
c_1\; i \varepsilon_{\mu \nu \rho \sigma} \la{\hat{\alpha}_L}^{\mu} {\hat{\alpha}_L}^{\nu} {\hat{\alpha}_L}^{\rho} {\hat{\alpha}_R}^{\sigma}-{\hat{\alpha}_R}^{\mu} {\hat{\alpha}_R}^{\nu} {\hat{\alpha}_R}^{\rho} {\hat{\alpha}_L}^{\sigma}\ra
+c_2\; i \varepsilon_{\mu \nu \rho \sigma} \la{\hat{\alpha}_L}^{\mu} {\hat{\alpha}_R}^{\nu} {\hat{\alpha}_L}^{\rho} {\hat{\alpha}_R}^{\sigma}\ra\notag\\
&\quad+c_3\;\varepsilon_{\mu \nu \rho \sigma} \la {F_V}^{\mu \nu} ({\hat{\alpha}_L}^{\rho} {\hat{\alpha}_R}^{\sigma}-{\hat{\alpha}_R}^{\rho} {\hat{\alpha}_L}^{\sigma})\ra
+ \frac{c_4}{2}\varepsilon_{\mu \nu \rho \sigma} \la {\hat{F}_L}^{\mu \nu} ({\hat{\alpha}_L}^{\rho} {\hat{\alpha}_R}^{\sigma}-{\hat{\alpha}_R}^{\rho} {\hat{\alpha}_L}^{\sigma}) \notag\\ &-{\hat{F}_R}^{\mu \nu} ({\hat{\alpha}_R}^{\rho} {\hat{\alpha}_L}^{\sigma}-{\hat{\alpha}_L}^{\rho} {\hat{\alpha}_R}^{\sigma})\ra\big].
\end{align}
With the building blocks in this paper, the results are
\begin{align}\label{hlsano}
\mathcal{L}_{\text{H},A,4}=&\;
{\cal B}_1\, i \varepsilon^{\mu \nu \lambda \rho}\langle\hat{\alpha}_{\perp \mu} \hat{\alpha}_{\perp \nu} \hat{\alpha}_{\perp \lambda} \hat{\alpha}_{\parallel \rho}\rangle
+{\cal B}_2\,\varepsilon^{\mu \nu \lambda \rho}(\langle\hat{\alpha}_{\perp \mu} \hat{\alpha}_{\parallel \nu} \Vs_{\lambda \rho}\rangle+\langle\hat{\alpha}_{\perp \mu} \Vs_{\lambda \rho}\hat{\alpha}_{\parallel \nu} \rangle)\notag\\
&+{\cal B}_3\,\varepsilon^{\mu \nu \lambda \rho}(\langle\hat{\alpha}_{\perp \mu} \hat{\alpha}_{\parallel \nu} V_{\lambda \rho}\rangle+\langle\hat{\alpha}_{\perp \mu} V_{\lambda \rho}\hat{\alpha}_{\parallel \nu} \rangle)
+{\cal B}_4\, i \varepsilon^{\mu \nu \lambda \rho}\langle\hat{\alpha}_{\perp \mu} \hat{\alpha}_{\parallel \nu} \hat{\alpha}_{\parallel \lambda} \hat{\alpha}_{\parallel \rho}\rangle.
\end{align}
These two sets of building blocks have the following relations:
\begin{align}
&{\hat{\alpha}_L}^{\mu}={\hat{\alpha}_\parallel}^{\mu}-{\hat{\alpha}_\perp}^{\mu},\quad {\hat{\alpha}_R}^{\mu} ={\hat{\alpha}_\parallel}^{\mu}+{\hat{\alpha}_\perp}^{\mu},\quad \varepsilon_{\mu\nu\rho\sigma}{F_V}^{\mu \nu} = \frac{1}{2}\varepsilon_{\mu\nu\rho\sigma}V^{\mu \nu},\\
& \varepsilon_{\mu\nu\rho\sigma}{\hat{F}_L}^{\mu \nu} = \frac{1}{2}\varepsilon_{\mu\nu\rho\sigma}(\Vs^{\mu \nu} - \As^{\mu \nu}),\quad \varepsilon_{\mu\nu\rho\sigma}{\hat{F}_R}^{\mu \nu} = \frac{1}{2}\varepsilon_{\mu\nu\rho\sigma}(\Vs^{\mu \nu}+\As^{\mu \nu}).
\end{align}
The relations of the LECs are
\begin{align}
{\cal B}_1=\frac{N_c}{16\pi^2}(-4c_1+4c_2),\quad{\cal B}_2=-\frac{N_c}{16\pi^2}c_4,\quad {\cal B}_3=-\frac{N_c}{16\pi^2}c_3,\quad {\cal B}_4=\frac{N_c}{16\pi^2}(-4c_1-4c_2).
\end{align}

\subsection{NNLO}\label{nnlo}

The NNLO results for the chiral Lagrangians are presented in Tables \ref{tnnlo} and \ref{hnnlo} in the supplementary material due to their length. Note that in the vector-field representation, the number of the independent terms at NNLO is significantly larger than those in the other representations. The number of terms is too large to be presented here. This highlights an advantage of the tensor-field representation, which requires fewer terms compared to the vector-field representation. The numbers of these chiral Lagrangians are listed in Table \ref{numnn}.

\begin{table*}[!h]
\caption{\label{numnn}The numbers of the NNLO chiral Lagrangians with vector mesons in the tensor-field and the HLS representations. For the HLS representation, the terms that correspond directly to those in the pseudoscalar mesonic chiral Lagrangians \cite{bijnens_mesonic_1999,bijnens_anomalous_2002} are ignored.}
\begin{center}
\begin{tabular}{c|cccc}
\hline\hline
      Flavor     & \multicolumn{2}{c}{$SU(2)$} & \multicolumn{2}{c}{$SU(3)$} \\
\hline
  Representation & Tensor &        HLS         & Tensor &        HLS         \\
      Normal     &  393   &        407         &  847   &        878         \\
     Anomaly     &  409   &        173         &  1325  &        493         \\
      Total      &  882   &        580         &  2172  &        1371        \\
\hline\hline
\end{tabular}
\end{center}
\end{table*}



\section{Relations between HLS and tensor-field representations}\label{sec:equivalence}
Although there are several different representations for introducing vector mesons in ChPT, they are related to each other. As noted in the introduction, the relationship between the vector-field and tensor-field representations has been previously studied. Here, we briefly discuss the connection between the HLS and tensor-field representations. Strict relations depend on multi-vector-meson vertices, but we only address single-vector-meson vertices in the tensor-field representation.

The relationship between the HLS and tensor-field representations is established via an auxiliary field method \cite{tanabashi_formulations_1996}. In this approach, an additional term
\begin{align}
\frac{1}{2}\kappa^2\left\la \left(V_\mu - a_{\parallel\mu} - \frac{1}{\kappa} D^\nu W_{\nu\mu}\right) \left(V^\mu - a_{\parallel}^{\mu} - \frac{1}{\kappa} D_\lambda W^{\lambda\mu}\right) \right\ra \label{aterm}
\end{align}
is introduced to the chiral Lagrangian in the tensor-field representation. By integrating out the field $W^{\mu\nu}$, one can derive the Lagrangian in the HLS representation. Here, $\kappa$ is an arbitrary parameter.

In Ref.~\cite{tanabashi_chiral_1993}, operators for four-point vertices are omitted. Including these four-point vertices yields the complete results as
\begin{align}
\begin{aligned}
&a =\frac{{\kappa}^{2}}{2F_\pi^2},\,g=\frac{\sqrt{2} M_V}{\kappa},\,z_1=-\frac{(\kappa-\sqrt{2}F_V)^2}{4M_V^2},\,z_3=\frac{\kappa(\kappa-\sqrt{2}{F_V})}{2M_V^2},\,z_4=\frac{ \kappa(\kappa-2 \sqrt{2} {G_V})}{M_V^2},\\
&z_5=-\frac{ {\kappa}^{2}}{M_V^2},\,z_6=-\frac{(\kappa-\sqrt{2}{F_V} ) (\kappa-2\sqrt{2} {G_V})}{M_V^2},\,z_7=\frac{\kappa(\kappa-\sqrt{2} {F_V} )}{M_V^2},\\
&y_1=-y_2=-\frac{(\kappa-2\sqrt{2} {G_V})^{2}}{2M_V^2},\,y_3=-y_4=- \frac{ {\kappa}^{2}}{2M_V^2},\,y_6=-y_7=-\frac{ \kappa(\kappa-2 \sqrt{2} {G_V})}{M_V^2},
\end{aligned}\label{lecrh}
\end{align}
where $F_V$ and $G_V$ are the leading order LECs in the tensor-field representation, $z_i$ and $y_i$ are the NLO LECs in the HLS representation. To our knowledge, the $y_i$ terms ($i=1, \dots, 7$) are not included in Ref.~\cite{tanabashi_formulations_1996} and are introduced here for the first time. Eq.~\eqref{lecrh} shows that leading-order terms in the tensor-field representation correspond to the specific NLO terms in the HLS representation. However, extending this method to NLO in the tensor-field representation does not generate additional NLO terms in the HLS representation. This limitation stems from two factors. First, the auxiliary-field term in Eq.~\eqref{aterm} is not unique, as it includes only a quadratic term involving vector sources. At the higher orders, integrating additional terms with vector sources, and potentially other external sources, would be required -- a process that becomes increasingly complex. Second, the tensor-field representation incorporates only single-vector vertices, while the HLS Lagrangian allows for vertices with multiple vectors. Although Eq.~\eqref{lecrh} yields some NLO terms, these are linked to the leading-order LECs in the tensor-field representation, making them dependent rather than independent NLO terms. Thus, if strong equivalence between the tensor-field and HLS representations is to hold, the number of independent LECs should match in both representations. However, this equivalence is not evident in Eq.~\eqref{lecrh} at the NLO level. Despite this, the auxiliary-field method does ensure that each term in the tensor-field representation has a corresponding term in the HLS representation.

Certain terms in the HLS representation are not derived in Eq.~\eqref{lecrh}. We will discuss which terms correspond to
\begin{align}
\la \ahp_{\mu}\ahe^{\mu}\ra\la\ahp_{\nu}\ahe^{\nu}\ra, \quad \la \ahp_{\mu}\ahp_{\nu}\ra\la\ahe^{\mu}\ahe^{\nu}\ra, \quad \la (\ahp_{\mu}\ahe^{\mu} - \ahe_{\mu}\ahp^{\mu})\hat{\chi}_- \ra.
\end{align}
Their associated LECs are $y_{14}$, $y_{15}$, and $w_5$, respectively. The method is similar to that presented in Ref.~\cite{tanabashi_formulations_1996}, except that the derivation proceeds from the HLS representation to the tensor-field representation. To simplify the calculation, we only consider the Lagrangian terms that include the kinetic term and other relevant contributions, while omitting terms that are not pertinent to this analysis. The simplified Lagrangian is
\begin{align}
\mathcal{L}_{\text{HLS}} = & F_\sigma^2 \la \hat{\alpha}_{\parallel\mu} \ahp^{\mu} \ra - \frac{1}{2g^2} \la V_{\mu \nu} V^{\mu \nu} \ra + y_{14} \la \hat{\alpha}_{\parallel\mu}\ahe^{\mu} \ra \la \hat{\alpha}_{\parallel\nu}\ahe^{\nu} \ra + y_{15} \la \hat{\alpha}_{\parallel\mu}\hat{\alpha}_{\parallel\nu} \ra \la \ahe^{\mu} \ahe^{\nu} \ra \notag \\
& + w_{5} \la (\hat{\alpha}_{\parallel\mu}\ahe^{\mu} - \ahe^{\mu}\hat{\alpha}_{\parallel\mu}) \hat{\chim} \ra + \cdots,
\end{align}
where ``$\cdots$'' denotes irrelevant terms. By introducing an auxiliary field $W^{\mu\nu}$ into the Lagrangian, we obtain
\begin{align}
\mathcal{L'}_{\text{HLS}} = \mathcal{L}_{\text{HLS}} + \frac{1}{2g^2 t^2} \la (W_{\mu \nu} + t V_{\mu \nu})(W^{\mu \nu} + t V^{\mu \nu}) \ra,
\end{align}
where $t$ is an arbitrary parameter that does not alter the dynamics of the system. Integrating out the vector field, which is implicitly present in $K^{\mu}$ within the HLS representation, leads to the final expression:
\begin{align}
\mathcal{L}'_{\text{T}}=& -\frac{1}{g^4 t^2 F_\sigma^2}\la\nabla_{\rho} W^{\rho \mu}\nabla^{\sigma} W_{\sigma \mu}\ra+\frac{1}{2g^2t^2}\la W_{\mu\nu}W^{\mu\nu}\ra+\frac{1}{2g^2t}\la W_{\mu\nu}f_+^{\mu\nu}\ra+\frac{1}{4g^2t}i\la W_{\mu\nu}\comm*{u^{\mu}}{u^{\nu}}\ra \notag\\
&-\frac{w_{5}}{2g^2t F_\sigma^2}\la W_{\mu\nu}\comm*{u^{\mu}}{\nabla^{\nu}\chim}\ra-\frac{w_{5}}{4g^2t F_\sigma^2}\la W_{\mu\nu}\comm*{f_-^{\mu\nu}}{\chim}\ra +\frac{y_{15}}{4g^4 t^2 F_\sigma^4}\la u_{\mu} u_{\nu}\ra\la\nabla_{\rho} W^{\rho \mu}\nabla_{\sigma} W^{\sigma \nu}\ra\notag\\
&+\frac{y_{14}}{4g^4 t^2 F_\sigma^4}\la u_{\mu}\nabla_{\rho} W^{\rho \mu} \ra\la u_{\nu}\nabla_{\sigma} W^{\sigma \nu}\ra+\frac{2}{g^8 t^3 F_\sigma^4}i\la\nabla_{\rho} W^{\rho \mu} W_{\mu\nu}\nabla_{\sigma} W^{\sigma \nu}\ra\notag\\
&+\frac{w_{5}}{2g^4 t^2 F_\sigma^4}\la\nabla_{\rho}W^{\rho\mu}[W_{\mu\nu},u^{\nu}\chim-\chim u^{\nu}]\ra +\cdots,\label{astcls}
\end{align}
where ``$\cdots$" denotes the pseudoscalar terms and higher-order terms, which are not relevant to this discussion. The first six terms correspond to the following Lagrangian in the tensor-field representation \cite{ecker_chiral_1989,cirigliano_towards_2006}:
\begin{align}
\mathcal{L}_{\text{T}} = & -\frac{1}{2} \la \nabla_{\rho} W^{\rho \mu} \nabla^{\sigma} W_{\sigma \mu} \ra + \frac{M_V^2}{4} \la W_{\mu\nu} W^{\mu\nu} \ra + \frac{1}{2\sqrt{2}} F_V \la W_{\mu\nu} f_+^{\mu\nu} \ra + \frac{1}{2\sqrt{2}} G_V i \la W_{\mu\nu} \comm*{u^{\mu}}{u^{\nu}} \ra \notag \\
& + D_{10} \la W_{\mu\nu} \comm*{u^{\mu}}{\nabla^{\nu} \chim} \ra + D_{20} \la W_{\mu\nu} \comm*{f_-^{\mu\nu}}{\chim} \ra, \label{lt}
\end{align}
where $D_{10}$ and $D_{20}$ are the NLO LECs in the tensor-field representation (see Table 1 in Ref.~\cite{cirigliano_towards_2006}). By comparing Eqs.~\eqref{astcls} and \eqref{lt}, the following relations are obtained:
\begin{align}
t = \frac{\sqrt{2}}{g^2 F_\sigma}, \quad M_V = g F_\sigma, \quad F_V = F_\sigma, \quad G_V = \frac{F_\sigma}{2}.\label{eq:t}
\end{align}
These relations are consistent with those found in Refs.~\cite{ecker_chiral_1989, tanabashi_formulations_1996}. The KSRF I relation $F_V = 2G_V$ is also derived from these. In Eq.~\eqref{astcls}, $y_{14}$ and $y_{15}$ are linked to $\la u_{\mu} \nabla_{\rho} W^{\rho \mu} \ra \la u_{\nu} \nabla_{\sigma} W^{\sigma \nu} \ra$ and $\la u_{\mu} u_{\nu} \ra \la \nabla_{\rho} W^{\rho \mu} \nabla_{\sigma} W^{\sigma \nu} \ra$, respectively, indicating that these terms correspond to two-vector vertices. The term $w_5$ is associated with $\la W_{\mu\nu} \comm*{u^{\mu}}{\nabla^{\nu} \chim} \ra$, $\la W_{\mu\nu} \comm*{f_-^{\mu\nu}}{\chim} \ra$, and $\la \nabla_{\rho} W^{\rho\mu} [W_{\mu\nu}, u^{\nu} \chim - \chim u^{\nu}] \ra$. Notably, the last term also involves two-vector vertices. The $w_5$ term thus incorporates both one- and two-vector vertices. This general observation implies that a single term in the HLS representation may correspond to multiple terms in the tensor-field representation, which could involve varying numbers of vector fields. This example highlights that certain terms in the HLS representation are associated with multi-vector vertices in the tensor-field representation. This conclusion is evident from the HLS Lagrangian, since one $\hat{a}_\parallel$ corresponds to one vector field. Therefore, achieving full equivalence requires considering all possible numbers of vector vertices, though this remains a challenging task at present.

\section{Conclusions and discussions}\label{sec:conclusions}

In this work, the chiral Lagrangians with vector meson nonet are constructed to the NLO in the vector-field representation, as well as the NNLO in tensor-field and HLS representations, considering both the normal and anomalous parts. Additional constraint conditions of octet are derived in the tensor-field and HLS representations at the NLO. With these new constraints, we identify that some terms in the literature are not linearly independent of the others and should be removed. Finally, we discuss the equivalence between the HLS and tensor-field representations. Achieving full equivalence requires consideration of all numbers of vector vertices.

Here we just consider single-vector vertices, while chiral effective Lagrangians involving two or more vector fields could be more interesting for investigating the vector meson masses, vector-vector meson interactions, electromagnetic properties and so on. It is interesting to extend our discussion to these cases. Extending this discussion to include all cases is beyond the scope of this paper and will be explored in future research.

From the above results, it can be seen that different representations contain different terms and LECs. The equivalence has been reviewed in the introduction. One could adopt any favorite representation. For example, Ref. \cite{ecker_chiral_1989} calculated the electromagnetic form factor of the pion in both vector-field and tensor-field representations. Although the vector-field representation gives more LECs, both representations give equivalent results. Of course, a one-to-one correspondence between LECs in different representations is generally absent, see the discussion below Eq. \eqref{eq:t}. In practice, different representations can be mixed in one problem. For example, the tensor-field representation for vector mesons and the vector-field representation for $J/\psi$ are chosen in Refs.~\cite{Chen:2014yta,Yan:2023nqz,Zhang:2025vbf}, where $J/\psi$ can be regarded as a vector meson singlet. In order to avoid complications, a representation that contains fewer LECs at a given order should simplify the calculation.

Although the higher-order chiral Lagrangians contain many terms, only a subset is relevant for specific problems. For instance, when studying pseudoscalar mesons using the vector mesonic chiral Lagrangian, only a few terms are significant, as discussed in Ref. \cite{Guo:2020cmv}. Nowadays, some complex computations can be efficiently handled using computer algebra systems. Tedious mathematical calculations become manageable. The higher-order contributions to the properties of vector mesons, such as their masses, electromagnetic properties, and interactions with pseudoscalar mesons, can be studied.

\section*{Acknowledgements}

We thank Professor Yong-Liang Ma for helpful discussions. This project is supported by the Guangxi Science Foundation under Grants No. 2025GXNSFAA069930, the Guangxi Science and Technology Innovation Platform Program (Leitai Action Plan, Grant No. Guike LT2600640026), Guangxi Key R\&D Program (Guangxi Funeng Action Plan, Grant No. Guike FN2504240040), the ``Guangxi Highland of Innovation Talents'' Program, the Natural Science Foundation of Sichuan Province under Grant No. 2026NSFSC0745 and the Doctoral Research Initiation Project of
China West Normal University under Grants No. 25KE038.
%

}
\bibliography{refs.bib}

\newpage

{
\title{Higher-order chiral Lagrangians with vector meson nonet in different representations: supplementary
material}
\author{Wei Guo$^{1,2}$}
\email{wguo@cwnu.edu.cn}
\author{Qin-He Yang$^{1,3}$}
\author{Shao-Zhou Jiang$^{1}$}
\email[Corresponding author.]{jsz@gxu.edu.cn}
\affiliation{$^{1}$Guangxi Key Laboratory for Relativistic Astrophysics, School of Physical Science and Technology, Guangxi University, Nanning 530004, China\\
$^{2}$School of Physics and Astronomy, China West Normal University, Nanchong 637009, China\\
$^{3}$School of Physics and Electronics, Hunan University, Changsha 410082, China}
\date{\today}

\begin{abstract}
This document provides the explicit form of the chiral Lagrangians for vector mesons at next-to-leading order (NLO) and next-to-next-to-leading order (NNLO) in different representations.

\end{abstract}
\maketitle


\section{NLO chiral Lagrangians in vector-field representation}

\begin{longtable}{lcclcc}
\caption{The NLO chiral Lagrangians with vector mesons in the vector-field representation.}\label{vnlo}\\
\hline\hline $P_n$ & $SU(2)$ & $SU(3)$ & $P_n$ & $SU(2)$ & $SU(3)$\\
\hline\endfirsthead

\hline\hline $P_n$ & $SU(2)$ & $SU(3)$ & $P_n$ & $SU(2)$ & $SU(3)$\\
\hline\endhead

\hline\hline 
\endfoot

\hline\endlastfoot

$i\la V^{\mu}u_{\mu}u^{\nu}u^{\lambda}f_{-\nu\lambda}\ra+\mathrm{H.c.}$ & 1 & 1 & $\la V^{\mu}u_{\mu}u^{\nu}\chim u_{\nu}\ra+\mathrm{H.c.}$ &  & 117  \\
$i\la V^{\mu}u^{\nu}u_{\mu}u^{\lambda}f_{-\nu\lambda}\ra+\mathrm{H.c.}$ & 2 & 2 & $\la V^{\mu}u^{\nu}u_{\mu}\chim u_{\nu}\ra+\mathrm{H.c.}$ &  & 118  \\
$i\la V^{\mu}u^{\nu}u_{\nu}u^{\lambda}f_{-\mu\lambda}\ra+\mathrm{H.c.}$ & 3 & 3 & $\la V^{\mu}\ra\la u_{\mu}u^{\nu}u_{\nu}\chim\ra+\mathrm{H.c.}$ &  & 119  \\
$i\la V^{\mu}u^{\nu}u^{\lambda}u_{\mu}f_{-\nu\lambda}\ra+\mathrm{H.c.}$ &  & 4 & $\la u^{\mu}\chim\ra\la V^{\nu}u_{\nu}u_{\mu}\ra+\mathrm{H.c.}$ &  & 120  \\
$i\la V^{\mu}u^{\nu}u^{\lambda}u_{\nu}f_{-\mu\lambda}\ra+\mathrm{H.c.}$ & 4 & 5 & $\la\chim\ra\la V^{\mu}u_{\mu}u^{\nu}u_{\nu}\ra+\mathrm{H.c.}$ &  & 121  \\
$i\la V^{\mu}u^{\nu}u^{\lambda}u_{\lambda}f_{-\mu\nu}\ra+\mathrm{H.c.}$ &  & 6 & $\la V^{\mu}\nabla^{\nu}f_{-\mu\nu}\chim\ra+\mathrm{H.c.}$ & 72 & 122  \\
$i\la V^{\mu}u_{\mu}u^{\nu}{f_{-\nu}}^{\lambda}u_{\lambda}\ra+\mathrm{H.c.}$ & 5 & 7 & $i\la V^{\mu}{f_{+\mu}}^{\nu}u_{\nu}\chim\ra+\mathrm{H.c.}$ & 73 & 123  \\
$i\la V^{\mu}u^{\nu}u_{\mu}{f_{-\nu}}^{\lambda}u_{\lambda}\ra+\mathrm{H.c.}$ &  & 8 & $i\la V^{\mu}{f_{+\mu}}^{\nu}\chim u_{\nu}\ra+\mathrm{H.c.}$ & 74 & 124  \\
\hline
\end{longtable}

\section{NNLO chiral Lagrangians in tensor-field representation}
\begin{longtable}{lcclcc}
\caption{The NNLO chiral Lagrangians with vector meson in the tensor-field representation}\label{tnnlo}\\
\hline\hline $P_n$ & $SU(2)$ & $SU(3)$ & $P_n$ & $SU(2)$ & $SU(3)$\\
\hline\endfirsthead

\hline\hline $P_n$ & $SU(2)$ & $SU(3)$ & $P_n$ & $SU(2)$ & $SU(3)$\\
\hline\endhead

\hline\hline 
\endfoot

\hline\endlastfoot

$i\la W^{\mu\nu}u_{\mu}u_{\nu}u^{\lambda}u_{\lambda}u^{\rho}u_{\rho}\ra+\mathrm{H.c.}$ & 1 & 1 & $i\varepsilon^{\mu\nu\lambda\rho}\la{W_{\mu}}^{\sigma}u_{\sigma}u_{\nu}\ra\la u^{\delta}u_{\delta}f_{-\lambda\rho}\ra+\mathrm{H.c.}$ &  & 1104  \\
$i\la W^{\mu\nu}u_{\mu}u_{\nu}u^{\lambda}u^{\rho}u_{\lambda}u_{\rho}\ra+\mathrm{H.c.}$ & 2 & 2 & $i\varepsilon^{\mu\nu\lambda\rho}\la{W_{\mu}}^{\sigma}u^{\delta}\ra\la u_{\nu}u_{\sigma}u_{\delta}f_{-\lambda\rho}\ra+\mathrm{H.c.}$ &  & 1105  \\
$i\la W^{\mu\nu}u_{\mu}u_{\nu}u^{\lambda}u^{\rho}u_{\rho}u_{\lambda}\ra+\mathrm{H.c.}$ &  & 3 & $i\varepsilon^{\mu\nu\lambda\rho}\la{W_{\mu}}^{\sigma}u^{\delta}u_{\nu}\ra\la u_{\sigma}u_{\delta}f_{-\lambda\rho}\ra+\mathrm{H.c.}$ &  & 1106  \\
$i\la W^{\mu\nu}u_{\mu}u^{\lambda}u_{\nu}u_{\lambda}u^{\rho}u_{\rho}\ra+\mathrm{H.c.}$ & 3 & 4 & $i\varepsilon^{\mu\nu\lambda\rho}\la{W_{\mu}}^{\sigma}u_{\nu}\ra\la u_{\sigma}u_{\lambda}u^{\delta}f_{-\rho\delta}\ra+\mathrm{H.c.}$ &  & 1107  \\
$i\la W^{\mu\nu}u_{\mu}u^{\lambda}u_{\nu}u^{\rho}u_{\lambda}u_{\rho}\ra+\mathrm{H.c.}$ & 4 & 5 & $i\varepsilon^{\mu\nu\lambda\rho}\la{W_{\mu}}^{\sigma}u_{\nu}\ra\la u_{\sigma}u^{\delta}u_{\delta}f_{-\lambda\rho}\ra+\mathrm{H.c.}$ &  & 1108  \\
$i\la W^{\mu\nu}u_{\mu}u^{\lambda}u_{\nu}u^{\rho}u_{\rho}u_{\lambda}\ra+\mathrm{H.c.}$ &  & 6 & $i\varepsilon^{\mu\nu\lambda\rho}\la u^{\sigma}f_{-\mu\nu}\ra\la{W_{\lambda}}^{\delta}u_{\rho}u_{\delta}u_{\sigma}\ra+\mathrm{H.c.}$ & 504 & 1109  \\
$i\la W^{\mu\nu}u_{\mu}u^{\lambda}u_{\lambda}u_{\nu}u^{\rho}u_{\rho}\ra+\mathrm{H.c.}$ &  & 7 & $i\varepsilon^{\mu\nu\lambda\rho}\la{W_{\mu}}^{\sigma}u_{\nu}\ra\la u_{\lambda}u^{\delta}u_{\delta}f_{-\rho\sigma}\ra+\mathrm{H.c.}$ & 505 &  \\
$i\la W^{\mu\nu}u_{\mu}u^{\lambda}u_{\lambda}u^{\rho}u_{\nu}u_{\rho}\ra+\mathrm{H.c.}$ &  & 8 & $i\varepsilon^{\mu\nu\lambda\rho}\la u^{\sigma}{f_{-\mu}}^{\delta}\ra\la W_{\nu\delta}u_{\lambda}u_{\rho}u_{\sigma}\ra+\mathrm{H.c.}$ & 506 &  \\
\hline
\end{longtable}

\section{ NNLO chiral Lagrangians in HLS representation}
\begin{longtable}{@{}l@{}c@{ }c@{ }c@{ }l@{}c@{ }c@{ }c}
\caption{The NNLO chiral Lagrangians with vector mesons in the HLS representation. Here, $n$ denotes a cumulative index.}
\label{hnnlo}\\
\hline\hline $P_n$  & $SU(2)$ & $SU(3)$ & $P_n$  & $SU(2)$ & $SU(3)$\\
\hline\endfirsthead

\hline\hline $P_n$ & $SU(2)$ & $SU(3)$ & $P_n$  & $SU(2)$ & $SU(3)$\\
\hline\endhead

\hline\hline 
\endfoot

\hline\endlastfoot

$i\la{\hat{a}_{\perp}}^{\mu}\hat{a}_{\perp\mu}{\hat{a}_{\perp}}^{\nu}{\hat{a}_{\perp}}^{\lambda}V_{\nu\lambda}\ra+\mathrm{H.c.}$ & 1 & 1 & $\la{\hat{a}_{\perp}}^{\mu}{\hat{a}_{\perp}}^{\nu}\hat{a}_{\parallel\mu}{\hat{a}_{\parallel}}^{\lambda}\hat{a}_{\parallel\lambda}\hat{a}_{\parallel\nu}\ra$ & 343 & 687  \\
$i\la{\hat{a}_{\perp}}^{\mu}{\hat{a}_{\perp}}^{\nu}\hat{a}_{\perp\mu}{\hat{a}_{\perp}}^{\lambda}V_{\nu\lambda}\ra+\mathrm{H.c.}$ & 2 & 2 & $\la{\hat{a}_{\perp}}^{\mu}{\hat{a}_{\perp}}^{\nu}\hat{a}_{\parallel\nu}\hat{a}_{\parallel\mu}{\hat{a}_{\parallel}}^{\lambda}\hat{a}_{\parallel\lambda}\ra+\mathrm{H.c.}$ & 344 & 688  \\
$i\la{\hat{a}_{\perp}}^{\mu}{\hat{a}_{\perp}}^{\nu}\hat{a}_{\perp\nu}{\hat{a}_{\perp}}^{\lambda}V_{\mu\lambda}\ra$ &  & 3 & $\la{\hat{a}_{\perp}}^{\mu}{\hat{a}_{\perp}}^{\nu}\hat{a}_{\parallel\nu}{\hat{a}_{\parallel}}^{\lambda}\hat{a}_{\parallel\mu}\hat{a}_{\parallel\lambda}\ra+\mathrm{H.c.}$ & 345 & 689  \\
$i\la{\hat{a}_{\perp}}^{\mu}{\hat{a}_{\perp}}^{\nu}{\hat{a}_{\perp}}^{\lambda}\hat{a}_{\perp\mu}V_{\nu\lambda}\ra$ &  & 4 & $\la{\hat{a}_{\perp}}^{\mu}{\hat{a}_{\perp}}^{\nu}\hat{a}_{\parallel\nu}{\hat{a}_{\parallel}}^{\lambda}\hat{a}_{\parallel\lambda}\hat{a}_{\parallel\mu}\ra$ &  & 690  \\
$i\la{\hat{a}_{\perp}}^{\mu}\hat{a}_{\perp\mu}\ra\la{\hat{a}_{\perp}}^{\nu}{\hat{a}_{\perp}}^{\lambda}V_{\nu\lambda}\ra$ &  & 5 & $\la{\hat{a}_{\perp}}^{\mu}{\hat{a}_{\perp}}^{\nu}{\hat{a}_{\parallel}}^{\lambda}\hat{a}_{\parallel\mu}\hat{a}_{\parallel\nu}\hat{a}_{\parallel\lambda}\ra$ & 346 & 691  \\
$i\la{\hat{a}_{\perp}}^{\mu}V^{\nu\lambda}\ra\la\hat{a}_{\perp\mu}\hat{a}_{\perp\nu}\hat{a}_{\perp\lambda}\ra$ &  & 6 & $\la{\hat{a}_{\perp}}^{\mu}{\hat{a}_{\perp}}^{\nu}{\hat{a}_{\parallel}}^{\lambda}\hat{a}_{\parallel\nu}\hat{a}_{\parallel\mu}\hat{a}_{\parallel\lambda}\ra$ &  & 692  \\
\hline
\end{longtable}



\end{document}